\documentclass[11pt, oneside]{Thesis} 

\graphicspath{{Pictures/}} 

\usepackage[square, comma, sort&compress]{natbib} 
\hypersetup{urlcolor=blue, colorlinks=true} 

\title{Understanding user search processes across varying cognitive levels} 
\usepackage{amssymb}
\usepackage{arydshln}
\usepackage{enumitem}
\usepackage{soul}
\usepackage{pifont}
\begin{document}

\frontmatter 

\setstretch{1.3} 

\fancyhead{} 
\rhead{\thepage} 
\lhead{} 

\pagestyle{fancy} 

\newcommand{\HRule}{\rule{\linewidth}{0.5mm}} 

\hypersetup{pdftitle={\ttitle}}
\hypersetup{pdfsubject=\subjectname}
\hypersetup{pdfauthor=\authornames}
\hypersetup{pdfkeywords=\keywordnames}


\begin{titlepage}
\begin{center}

\textsc{\LARGE \univname}\\[1.5cm] 
\textsc{\Large Masters Thesis}\\[0.5cm] 

\HRule \\[0.4cm] 
{\huge \bfseries \ttitle}\\[0.4cm] 
\HRule \\[1.5cm] 
 
\begin{minipage}{0.4\textwidth}
\begin{flushleft} \large
\emph{Author:}\\
\href{}{\authornames} 
\end{flushleft}
\end{minipage}
\begin{minipage}{0.4\textwidth}
\begin{flushright} \large
\emph{Examiners:} \\
\href{}{\examname}  
\end{flushright}
\end{minipage}\\[1cm]
\begin{minipage}{0.8\textwidth}
\begin{flushright} \large
\emph{Supervisor:} \\
\href{}{\supname} 
\end{flushright}
\end{minipage}\\[2cm]\large \textit{A thesis submitted in fulfillment of the requirements\\ for the degree of \degreename}\\[0.3cm] 
\textit{in the}\\[0.4cm]

\deptname\\[2cm] 
 
{\large \today}\\[1cm] 
\includegraphics[width=6cm]{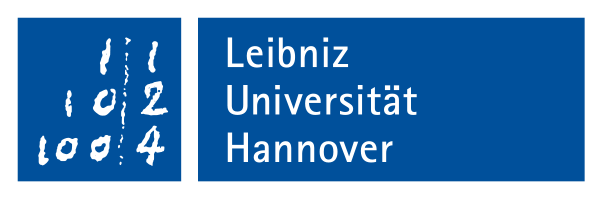} 
 
\vfill
\end{center}

\end{titlepage}


\Declaration{

\addtocontents{toc}{\vspace{1em}} 

I, \authornames, declare that this thesis titled, `\ttitle' and the work presented in it is my own. I confirm that this work submitted for assessment is my own and is
  expressed in my own words. Any uses made within it of the works of
  other authors in any form (e.g., ideas, equations, figures, text,
  tables, programs) are properly acknowledged at any point of their
  use. A list of the references employed is included.

 \vspace{2cm} 
Signed:\\
\rule[1em]{25em}{0.5pt} 
 
Date:\\
\rule[1em]{25em}{0.5pt} 
}

\clearpage 









\addtotoc{Abstract} 


 {\huge{\textit{Abstract}} \par}{\addtocontents{toc}{\vspace{1em}} }

Web is often used for finding information and with a learning intention. In this thesis, we propose a study to investigate the process of learning online across varying cognitive learning levels using crowd-sourced participants. Our aim was to study the impact of cognitive learning levels on search as well as increase in knowledge. We present 150 participants with 6 search tasks for varying cognitive levels and collect user interactions and submitted answers as user data. We present quantitative analysis of user data which shows that the outcome for all cognitive levels is learning by quantifying it as calculated knowledge gain. Further, we also investigate the impact of cognitive learning level on user interaction and knowledge gain with the help of user data. We demonstrate that the cognitive learning level of search session has a significant impact on user's search behavior as well as on knowledge that is gained. Further, we establish a pattern in which the search behavior changes across cognitive learning levels where the least complex search task has minimum number of user interactions and most complex search task has maximum user interactions. With this observation, we were able to demonstrate a relation between a learner's search behavior and \citeauthor{krathwohl2002revision}'s revised Bloom's taxonomic structure of cognitive processes. The findings of this thesis intend to provide a significant work to bridge the relation between search, learning, and user.
%

\clearpage 


\setstretch{1.3} 

\acknowledgements{\addtocontents{toc}{\vspace{1em}} 

I would first like to thank my thesis advisor Prof. Dr. Avishek Anand for allowing me to work on an interesting topic and providing constant guidance not only during the duration of the thesis but also throughout my Masters course. 

I would like to show an immense gratitude for my supervisor Dr. Ujwal Gadiraju who was always available to solve my doubts. His confidence in my work was most motivational. He was always available to discuss various approaches and steer me in the right direction. During the stressful periods of thesis, he made himself available even during his non-working days for which I will forever be grateful.

I would also like to thank Prof. Dr. Wolfgang Nejdl for giving me a working space in L3S Research Center during the entire duration for my masters and for funding the experimental work carried out in thesis.

I would like to thank my friends who were always present to support me emotionally and help me cope with the ups and downs of life during my entire masters course.

Finally, I would like to thank my family, especially my grandfather who taught me the importance of education since I was a child, my loving mother who always saw to it that I fed myself on time and lastly, my sister who was always awake when I needed a break.
}
\clearpage 


\pagestyle{fancy} 

\lhead{\emph{Contents}} 
\tableofcontents 

\lhead{\emph{List of Figures}} 
\listoffigures 

\lhead{\emph{List of Tables}} 
\listoftables 


\clearpage 

\setstretch{1.5} 

\lhead{\emph{Abbreviations}} 
\listofsymbols{ll} 
{

\textbf{DQ} & \textbf{D}istinct \textbf{Q}ueries \\
\textbf{IR} & \textbf{I}nformation \textbf{R}etrieval \\
\textbf{KG} & \textbf{K}nowledge \textbf{G}ain \\
\textbf{QL} & \textbf{Q}uery \textbf{L}ength \\
\textbf{RQ} & \textbf{R}esearch \textbf{Q}uestion \\
\textbf{SAL} & \textbf{S}earch \textbf{A}s \textbf{L}earning \\
\textbf{SERP} & \textbf{S}earch \textbf{E}ngine \textbf{R}esource \textbf{P}age \\
\textbf{URL} & \textbf{U}niform \textbf{R}esource \textbf{L}ocator \\
\textbf{UT} & \textbf{U}nique \textbf{T}erms \\

}

\mainmatter 

\pagestyle{fancy} 



\chapter{Introduction} 

\label{Chapter1} 

\lhead{Chapter 1. \emph{Introduction}} 
One of the most common and frequent usage of web is to find information. Whether it is for finding answers for a research question or to determine who won the best actor award, search is ingrained in our lives. While it seems that information seeking type searches are most common according to classic information retrieval, \citep{broder2002taxonomy} explains how not all search queries are for information need. He categorizes the search queries broadly into three forms - (i) navigational - queries that point to a particular domain or website, (ii) informational - queries that seek information, and (iii) transactional - queries with an intent of completing a transaction. While \citeauthor{broder2002taxonomy}'s categorization revises basic IR model, it does not help in identifying queries with learning needs, especially over different cognitive learning levels during the web search.

Most of the web search engines today are designed for satisfying either domain-specific search or an individual's look-up tasks. This optimization leads to using the extraordinary knowledge resource that is web search as a means to satiate immediate information need rather than as a learning tool. However, in order to design search engines that support complex search actions which are by-products of learning, it is important to understand and recognize these search actions first. Therefore, the focus of this thesis would be on understanding these search actions and to determine if it is possible to automatically recognize them. Hence, allowing us to distinguish learning related search actions from other types so as to optimize the search engines accordingly to provide the best results.
\section{Search as Learning}

Theoreticians like  Piaget and Vygotsky say that learning process depends upon existing knowledge on which factoids are built. This collaborates with how revised Bloom's taxonomy\citep{krathwohl2002revision} is built where each higher cognitive level in learning is based on lower ones. As discussed, even though web contains extraordinary resources for knowledge, the act of learning from this knowledge is much more than mere look-up of information and memorization of it\citep{Eickhoff2017}. Information look-up tasks are often simple and may get over in a small search session. However, in learning scenarios, the conditions are often reversed. The queries can be long and the search tasks required until entire learning process ends may span over many sessions. In the report \citep{collins2017search} which discusses Dagstuhl Seminar 17092 and consists of SAL topics spread over subjects like interactive IR, psychology, education and system-oriented IR, Yiqun Liu mentions that most of the discussion in SAL can be boiled down to questions like ``How can we model user's cognitive states?", ``Does a user's cognitive state affect its search behavior?" and ``What can be the consequences of a better search engine?". Liu mentions these questions in the context of SAL studies helping current search enginesni face challenges while serving exploratory search queries, multi-step search, complex search sessions which all are a result of learning while searching. In this thesis we will mainly focus on the question - ``Does a user's cognitive state affect its search behavior?".

In order to promote learning, a search session should be able to help in finding, understanding, analyzing, evaluating and creating documents that would contain information which would eventually provide answers to a complex question. This process is time-consuming and cognitively demanding and hence, requires an intelligent search system for the user. In order to make this intelligent search system, the search system in itself needs to learn how the user behaves while finding, understanding, analyzing, evaluating and creating documents that would answer the complex questions.

To create this intelligent search system which is able to perceive user's learning process, we need to first answer many questions like the what are the challenges involved in measuring knowledge gain while searching, how human interactions and searching are related, how human learning process takes place while searching, what is the context in which the learning occurs and how important the context is while considering learning process, etc. All these challenges make up Search as Learning. It raises questions like how do we define learning where general assumption in many cases is that seen is equal to understood.

People have become comfortable with searching the way it is offered today. Google, Bing, etc are not designed as a learning system, nevertheless, people use it for learning. However, in order to create a smarter technology which will help people get smarter, one needs to study how people get smart. Therefore, this thesis will focus on how learning occurs while searching.

In this thesis, we investigate learning theory in order to understand the information search. Our aim is to discover a framework  based on a learning theory to identify the relation between cognitive learning category and searcher. We intend to establish a relation between the two based on searcher's behavior and searcher's knowledge change.


\section{Research Questions and original Contributions}\label{sec:researchQues}
This thesis tries to understand user's search processes across cognitive levels of a taxonomic structure using distinct search patterns. In this thesis, a novel design for task setup is provided in order to give better insights to knowledge change in search sessions related to learning on the web as well on relation between user's search behavior and learning. The following research questions guide the overall direction and objectives of this master thesis:

\begin{enumerate}
    \item RQ1: How does a user's knowledge evolve in a search session online with respect to the varying cognitive learning levels?
    \item RQ2: How is search behavior impacted by cognitive learning level?
\end{enumerate}

In order to answer these research questions search tasks were developed and a unique crowd-sourcing experimental setup was designed such that each task tries to disassociate itself from other cognitive levels of taxonomy. The purpose of this approach comes with the hypothesis that by making each cognitive level discrete from other, the user will not be carrying any prior knowledge from lower level which will help in identifying user behavior correctly.
 
The following thesis is structured as follows. Chapter 2 discusses the relevant literature and related works. Chapter 3 provides a background for the thesis where the chosen taxonomic structure for cognitive levels is explored. Chapter 4 details about the original approach of this thesis. It is also here where the design of the experiments across various cognitive learning level are discussed. Chapter 5, the results chapter provides an analysis of the data collected and discusses whether this data is able to give a solution for the above research questions and finally Chapter 6 concludes the thesis and presents possible future directions for the work.


\chapter{Related Work} 

\label{Chapter2} 

\lhead{Chapter 2. \emph{Related Work}} 

Learning is not restricted to classrooms anymore. In addition, learning style in today's age is not constrained to a classroom setup of `teacher and students'. \citeauthor{wilson1981user} mentions that information need often refers to one's underlying motivation to seek the specific type of content. In his book \citep{schutz1973structures}, \citeauthor{schutz1973structures} discuss how every individual has his or her own view of the world around them, specific typification that are used to model and explain all the phenomena around them and when one encounters a problem which won't fit in their model, it requires more information and knowledge remodeling in order to solve the problem and fix the anomaly. Several literature talk about relationships between sense-making models and information seeking. \citep{dervin1983overview} views information seeking as a means to demolish the uncertainty between desired and observed scenarios. Further, \citep{dervin1998sense} reviews user's sense making approach by transforming user's conceptualization from noun based knowledge framework to verb based framework. Likewise, there have been discussions where strong relations between information seeking, knowledge and human cognition levels have been displayed. \citep{ingwersen1996cognitive}'s theory on text retrieval and cognitive framework as well as \citep{wilson1981user}'s problem solution model are a few of such examples.

However, the focus of this thesis is not only in information seeking but also in searching and learning of information. There exists an abundant amount of literature which emphasizes on relations between search, learning, and user or more specifically user behavior in the area of information science. \citep{Eryilmaz2013} explains how annotation in a collaborative environment affects learning of an individual in an online system through theoretical experiments. While \citep{stahl2000model} explores learning online by providing a knowledge building model in order to support collaborative learning, \citep{waters2006social} inspects learners' behavior in an online collaborative system.

In order to comprehend the interrelation between search, user, and learning; it is crucial to recognize the procedure of learning. Not only, is it essential to understand the process but it is also required to identify the different stages of learning. \citep{bloom1956taxonomy} developed taxonomic structure to encourage education at a deeper level as compared to mere fact recalling. The categories of taxonomic structure was viewed as learning levels. His motivation was to create thinkers in the world and in order to do so, he proposed six levels in the taxonomy where the levels on top of the structure were more abstract and required higher level of thinking and reasoning in contrast to lower levels. The six categories of Bloom's taxonomy from least abstract to most abstract are: \textit{Knowledge, Comprehension, Application, Analysis, Synthesis,} and \textit{Evaluation}. Bloom's taxonomy is not a perfect taxonomy to categorize and order learning levels. \citep{chan2002applying} supports this fact by arguing that there aren't any perfect educational taxonomies and many taxonomies have their weaknesses, including that of Bloom's. The original Bloom's taxonomy has loopholes and have been challenged by many. \citep{kunen1981levels} studies showed that while all the other categories of Bloom's taxonomy led to increase in memory, the \textit{Evaluation} level failed to do so. He questioned position and inclusion of \textit{Evaluation} in the educational taxonomy. \citep{krathwohl2002revision} argued the usability of Bloom's taxonomic structure in educational systems as educators are used to designing the learning objectives in a  ``subject-description'' format where subject would refer to subject matter of the content and description would include an explanation of how to deal with the content. He further illustrated that this ``subject-description'' format can also be viewed as a ``noun-verb'' pair. \citeauthor{krathwohl2002revision} modified the original taxonomy into 2-Dimensional where Knowledge formed one of the dimensions and cognitive processes of learning the knowledge formed the second dimension. The revision allowed evaluation for both, the learning outcomes as well as the cognitive process used by learner\citep{valcke2009supporting}. We will be adopting this revised version of Bloom's taxonomy for our study here and it will be used as a scripting guide for designing the tasks and measuring knowledge as well as user interactions at various cognitive levels.

\section{Task Modeling}
Designing search tasks is a difficult and time consuming problem as it asks for specialized knowledge. The modeling is further complicated by the abundance
of various research illustrating how variations in search tasks and search task properties can impact searcher behavior \citep{kelly2015development, wu2012grannies}.  Poorly designed search tasks can often lead to invalid results as users participate in unacceptable searches and depict inadmissible user behavior. This will hence, lead to wastage in money and time by complicating the analysis process. For example, it is not useful if we design a difficult search task where the learner can find the answer from the first Wikipedia page by firing a simple search query. While tasks can be classified in many ways, by its type - \textit{e.g., open, factual,
navigational, decision-making}, by its topic - \textit{e.g.,
difficulty, urgency, structure, stage}; in the current scenario we are interested in classifying the tasks by its complexity. In order to do so, we use \citeauthor{krathwohl2002revision}'s revised Bloom's taxonomy to classify the tasks into its six cognitive processes, much like \citep{kelly2015development, jansen2009using}. However, as \citep{jansen2009using} points out, designing tasks based on Anderson and Krathwohl's taxonomy is complicated as the categories of the taxonomy are not distinct from each other. It also implies that the revised Bloom's taxonomy has learning levels that overlap its boundaries with the it's immediate top and bottom levels. Therefore, it is crucial to design the search tasks properly such that it will call for its users to utilize the labeled cognitive process. In order to create the questions of search tasks for this study for each category, the design was heavily guided by previous literature \citep{ferguson2002using, lord2007moving} just like \citep{ kelly2015development, jansen2009using, ghosh2018searching}. However, in this work, we tweak the previous experimental setups which used search tasks and ended up not being able to find distinct user behaviors among different levels of cognitive processes. Most of these experiments did not establish the fact that the revised Bloom's taxonomy has overlapping levels while allocating tasks to learners. This would mean that if a learner performs tasks for two different levels, his behavior for other levels will be tainted as he is carrying knowledge from previous levels.

\section{Searching, Learning, and User-Interactions}

In order to solve the research questions, a quantitative study using crowd-sourced experimental design was conducted. In the recent past, we have seen a number of experiments carried out that revolve around finding distinct user characteristics if any among learners online. \citep{jansen2009using} used Bloom's taxonomy to design six search tasks on various topics and asked each participant of his experiment to solve the six searching problems. All these six searching problems were of same broader topic. His experimental analysis concluded an inverted curve relationship between cognitive learning level and searching difficulty as shown in Figure 2.1. He reasons that the the learners carry out search at higher cognitive level which are similar to those of at lower cognitive level as they already possess knowledge, presumably from middle levels and are more interested in mere verification of facts and theory. This explanation is justifiable as the revised Bloom's taxonomy is a continuous pyramid with no rigid distinction.

\begin{figure}
    \centering
    \includegraphics[width=\textwidth]{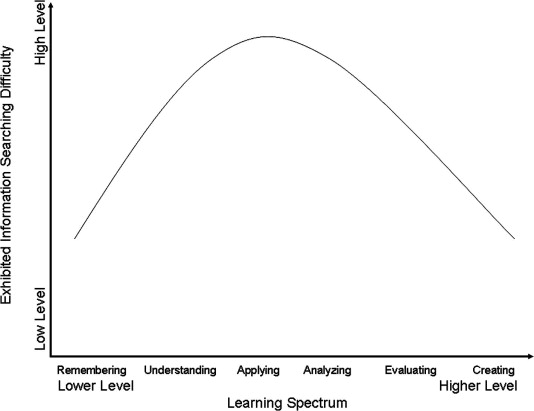}
    \caption{\citeauthor{jansen2009using}'s relation between cognitive processes of revised Bloom's taxonomy and search difficulty}
    \label{fig:jansen's_search_learning_relation}
\end{figure}

In the preliminary study \citep{wu2012grannies}, the authors created 20 search tasks for five cognitive levels of \citeauthor{krathwohl2002revision}'s taxonomy in four different domains and conducted a laboratory experiment with undergraduate students. The study includes questions for all levels of cognitive complexity except \textit{Apply}. Undergraduate students who participated in the study answered pre-task questionnaire, post-task questionnaire, and an exit interview. The domain of the five tasks that participants answered were assigned in a rotating manner. The results gathered from questionnaire, interviews, and task completion showed that while search interactions showed an increase with increase in level, the students marked the experienced task difficulty different to that of expected. \citep{kelly2015development} provides a more detailed insight for the above study. The study reported that there wasn't always a significant difference in search behaviors for tasks of mid-level cognitive complexity - \textit{Understand, Analyze, Evaluate,} in most of the cases, however, there were
significant differences between the lowest and highest tasks of \citeauthor{krathwohl2002revision}'s cognitive processes pyramid - \textit{Remember} and \textit{Create}.

In \citep{ghosh2018searching}, quantitative and qualitative analysis is provided to support the fact that there exists a relationship between search and learning. Much like previous literature, the researchers design tasks for undergraduate students according to \citeauthor{krathwohl2002revision}'s revised taxonomy. They set longer duration, spanning over weeks for the task and allow students to refer to online as well offline resources to gain knowledge as long as they log their exploration. \citeauthor{ghosh2018searching} designed four search tasks for different cognitive levels. \textit{Remember} and \textit{Understand} were clubbed into one search task which was of first or lowest order. The following three search tasks were for \textit{Apply, Analyze,} and \textit{Evaluate} respectively. The participants were provided with all four tasks in an hierarchical order of their complexity. The experimental data showed that while there were statistically significant results to support that learning was indeed an outcome from searching  there weren't always a significant difference in user's search behavior for tasks and cognitive complexity. The authors highlight the limitation of the design stating that since tasks were distributed to participants in an hierarchical order, it could have influenced their learning and hence, the results. 

According to common understanding, it is expected that the user-interactions and searching should increase as users climb higher in cognitive complexity of the search tasks. However, strong experimental proof is missing to support this theory. In this thesis, we try to discover this proof by tweaking the previous works and combining it with our original work. Further, we aim to support our theory by providing data for all six categories of of \citeauthor{krathwohl2002revision}'s revised taxonomy.

The framework of the experiments for this thesis can be viewed as extending the work of \citep{gadiraju2018analyzing, yu2018predicting} where the authors find knowledge gain of users in an informational search sessions. The rise in knowledge is measured by a 3 step set-up of pre-test, search session, and post-test where both pre-test and post-test questions are exactly same. The participating users were not aware of the fact that both pre-test and post-test are same, they were only made aware of the fact that the topic for both the tests is same. This experimental setup is replicated for the informational seeking task or also known as the task for \textit{Remember} level as it forms an elegant manner to measure information recall.

\chapter{Background} 

\label{Chapter3} 

\lhead{Chapter 3. \emph{Background}} 


\section{Bloom's Taxonomy}
Bloom's Taxonomy\citep{bloom1956taxonomy} was created by Benjamin Bloom in 1956. Benjamin Bloom provides a framework to categorize the levels of reasoning skills required in classroom like learning situations. Benjamin et al.'s taxonomy was designed in a manner to guide the educator in helping their students' learning progress. As an educator, the goal should be to move their learners higher in the taxonomy so the knowledge is progressed. This taxonomy broke the conventional education system where assessment of knowledge was based on recall of information. Bloom's taxonomy includes higher cognitive levels instead of just recall. There are six levels in the taxonomy. Each level requires a higher level of abstraction than the previous one from learners. The framework became a medium for facilitating the exchange of test items among faculty members of various universities\citep{krathwohl2002revision}. We will refer to this Bloom's taxonomy which was published in 1956 as \textit{Taxonomy of Educational Objectives: The Classification of Educational Goals. Handbook I: Cognitive Domain} as the ``original taxonomy" hereafter in this thesis.

\subsection{The Original Taxonomy}\label{sec:oriTaxonomy}

\begin{figure}
    \centering
    \includegraphics[width=\textwidth]{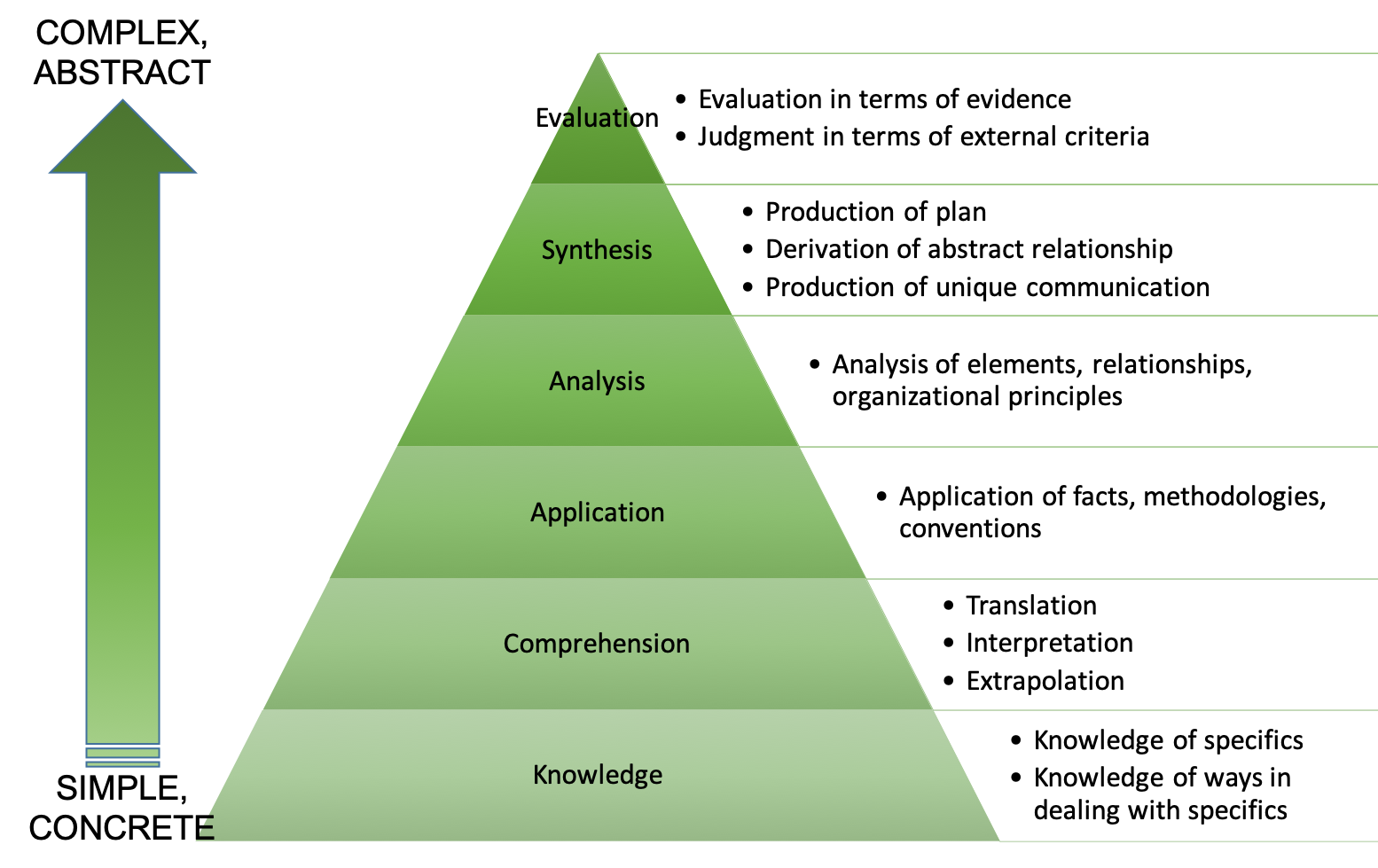}
    \caption{Original Bloom's taxonomy}
    \label{fig:original_taxonomy}
\end{figure}

Figure \ref{fig:original_taxonomy} describes original taxonomy as a pyramid structure where, the higher levels are more complex and abstract while lower levels are simple and concrete. The taxonomy contains six levels labeled as following: \textit{Knowledge, Comprehension, Application, Analysis, Synthesis, Evaluation}. The cumulative hierarchy of the taxonomy can allow us to assume that each simpler level was a prerequisite in order to master the more complex next level.
\subsubsection{Knowledge}
Bloom's taxonomy describes knowledge level of taxonomy as a level which is dedicated solely to test whether a learner has gained specific information. The tasks in knowledge level are memorization tasks. The tests at such a level can include memory of specifics as well as that of ideas. Typical words describing knowledge tests would be words like tell, list, label, name, etc.

\subsubsection{Comprehension}
Comprehension level of taxonomy require learners to understand the information and hence, push them beyond simple recall of information. It requires learners to interpret information. Words like describe, contrast, discuss, predict, etc. are used to make comprehension questions.

\subsubsection{Application}
Application level, as the name suggests require its learners to apply the knowledge. This implies to be able to solve a task by using information gained previously. Application type questions use words like complete, solve, examine, illustrate, show, etc.

\subsubsection{Analysis}
In this level, a learner is required to go beyond and have the ability to detect patterns that they can use to analyze a problem. Analysis questions can be formed by using words like analyze, explain, investigate, infer, etc.

\subsubsection{Synthesis}
Synthesis require a learner to create new predictions, plans or theories based on the facts at hand. This might require knowledge from multiple subjects and be able to synthesize this information from multiple subjects before formulating a conclusion. Questions that use words like invent, imagine, create, compose, etc. generally are synthesis questions.

\subsubsection{Evaluation}
Evaluation, the highest level of taxonomy expects learners to evaluate or judge information and conclude aspects like its value, bias, etc. When words like select, judge, debate, recommend, etc. are used, the question is generally an evaluation.

\subsection{Revision of Bloom's taxonomy}
\citep{krathwohl2002revision} mentions in his paper that objectives that describe intended learning outcomes are usually framed in terms of (i) some subject matter and (ii) description of what is to be done with or to the content. Hence, learning objectives can be viewed as a `noun-verb' pair where noun phrase is the subject matter content and verb phrase are the cognitive processes of learning. eg: learner will remember all the elements of periodic table has noun phrase as ``learner will" and verb phrase as ``remember all the elements of periodic table". With the noun-verb phrase description it is clear what is expected from learner, i.e, to \textit{remember} the given information.

\begin{figure}
    \centering
    \includegraphics[width=\textwidth]{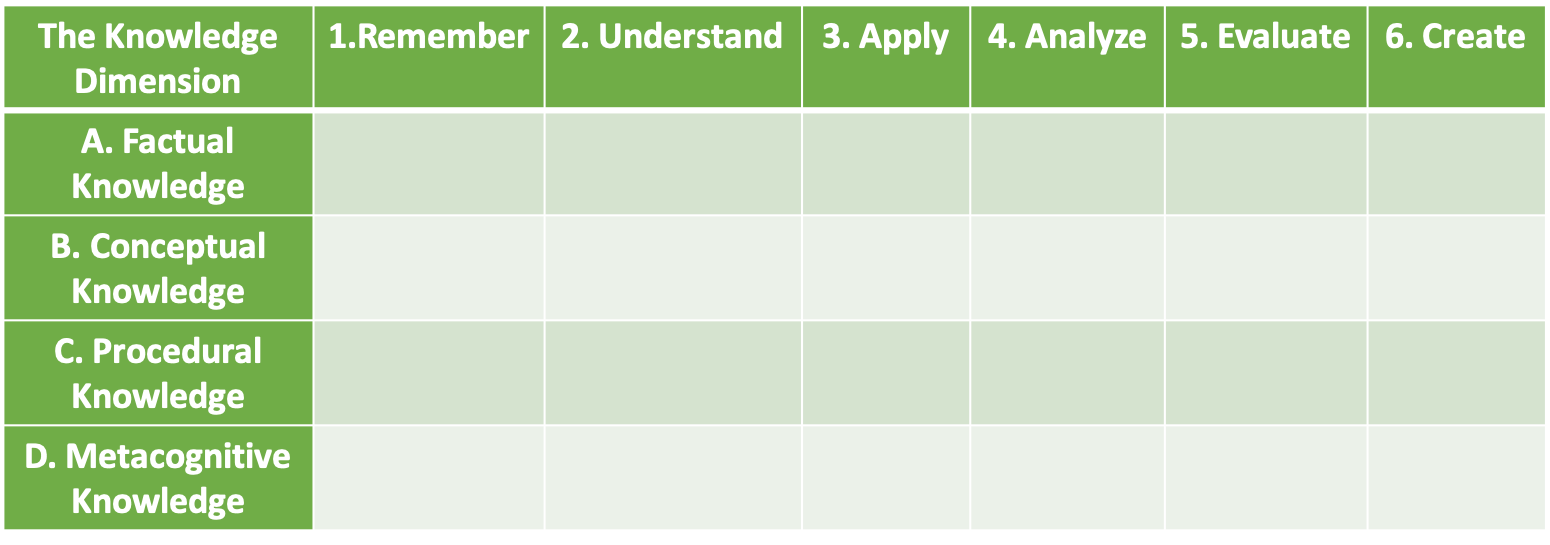}
    \caption{Two dimensions of revised Bloom's taxonomy}
    \label{fig:blooms_taxonomy2D}
\end{figure}

If we refer to the previous section \ref{sec:oriTaxonomy}, the original taxonomy had both noun and verb phrases included in its taxonomic structure. \textit{Knowledge} level had both noun as well as verb aspects. The verb aspect was how knowledge was defined by original taxonomy and noun aspect was the intent of \textit{knowledge}. Hence, we can see that the original taxonomy was uni-dimensional. In order to overcome this uni-dimensional drawback, \citeauthor{krathwohl2002revision} revised the original Bloom's taxonomy which will be referred to as \textit{Revised Bloom's taxonomy} from hereon in this thesis. In this revised Bloom's taxonomy, the noun provided the basis for Knowledge dimension and verb formed the basis for Cognitive process dimension. The new knowledge dimension contains four main categories namely, \textit{Factual Knowledge, Conceptual Knowledge, Procedural Knowledge,} and \textit{Metacognitive Knowledge}. The cognitive process dimension has six sub-categories namely, \textit{Remember, Understand, Apply, Analyze, Evaluate,} and \textit{Create}. Figure \ref{fig:blooms_taxonomy2D} shows how the revised bloom's taxonomy can form two dimensions in a tabular structure and hence, helps in categorization of objectives.

\begin{figure}
    \centering
    \includegraphics[width=\textwidth]{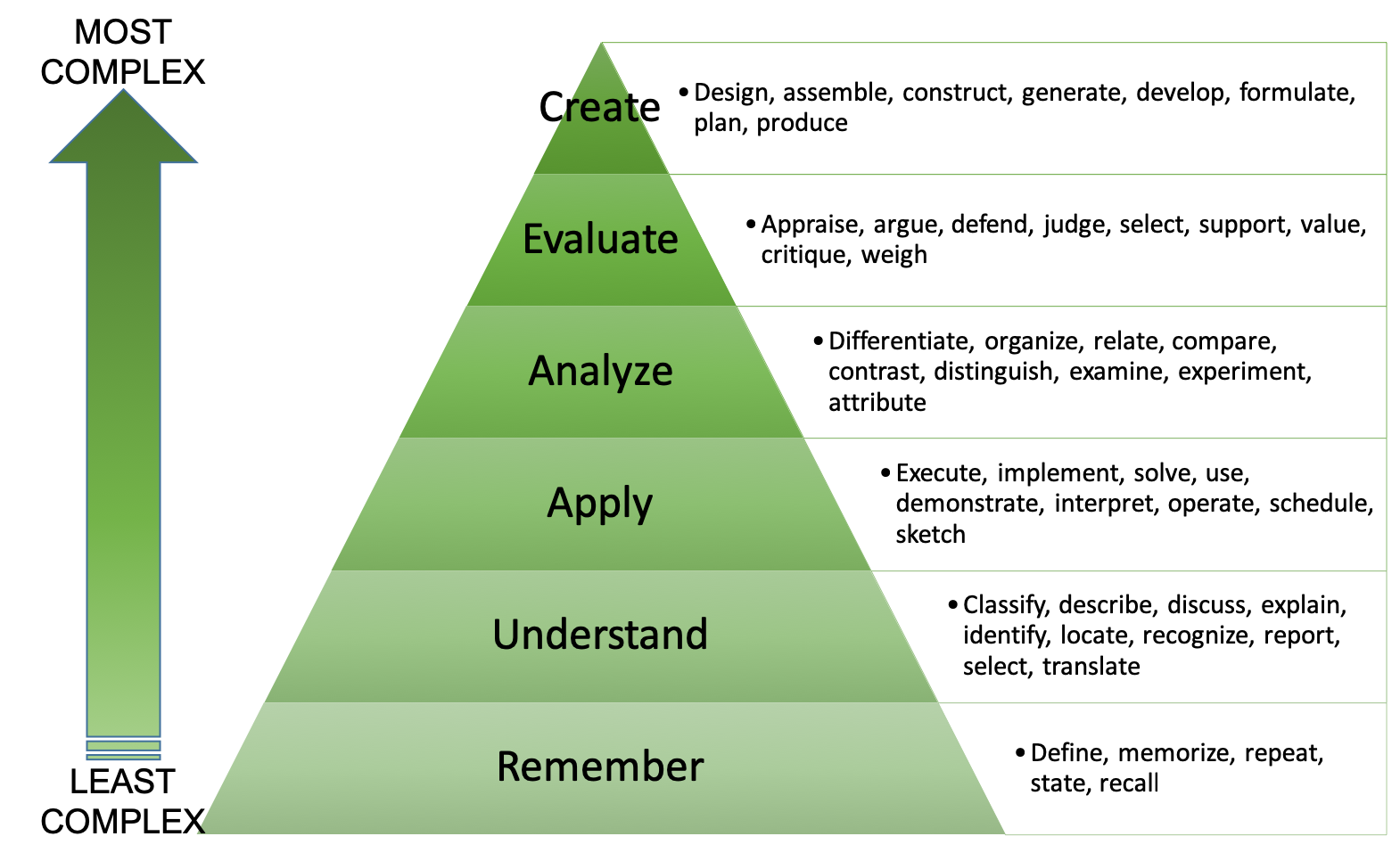}
    \caption{Cognitive processes of revised Bloom's taxonomy}
    \label{fig:blooms_taxonomy}
\end{figure}

The sub-categories of cognitive processes coincide with the categories of the original taxonomy. However, three categories from the original were renamed in revised version and two were swapped in the hierarchical structure. \textit{Knowledge} was renamed to \textit{Remember} to signify its ``verb-phrase". \textit{Comprehension} and \textit{Synthesis} were renamed to \textit{Understand} and \textit{Create} respectively. \textit{Application}, \textit{Analysis}, and \textit{Evaluation} were kept but titled in its verb form - \textit{Apply}, \textit{Analyze}, and \textit{Evaluate}. Finally, \textit{Evaluate} and \textit{Create}, namely \textit{Synthesis} and \textit{Evaluation} from original taxonomy swap complexity hierarchy in the revised version. \citeauthor{krathwohl2002revision} mentions that the revised version of  taxonomy, like the original is hierarchy where the cognitive process dimension vary in complexity. Figure \ref{fig:blooms_taxonomy} shows this hierarchical structure of cognitive process dimension along with words describing each dimension. These words describing each cognitive level can be used to define questions and tasks for the respective cognitive level. 

However, it is important to note that \citeauthor{krathwohl2002revision} also states, that hierarchy in revised Bloom's taxonomy is relaxed and that sub-categories of cognitive processes overlap with each other. The hypotheses in this thesis will be based on \citeauthor{krathwohl2002revision}'s statement that although there is hierarchical structure in the cognitive processes of revised Bloom's taxonomy, this is not a strict hierarchy. Understandably so, each cognitive process would have an overlap with its boundary processes as learning is a continuous process. 

\chapter{Approach} 

\label{Chapter4} 

\lhead{Chapter 4. \emph{Approach}} 

For this thesis, we proposed a unique experimental set-up to carry out a quantitative analysis on the crowd-sourced data accumulated by running different search sessions for varying cognitive stages of learning. The unique set-up tried to segregate every levels of the cognitive learning. Accumulated data from study consisted of logged user-interactions as well as submitted answers by users. We tried to determine the answers for the proposed research questions by examining the collected search behavior statistics of users. This chapter is divided into a motivation section \ref{sec:motivationDesign} where we discuss the motivation that led to the proposed study and our hypotheses, followed by a design section \ref{sec:design} which gives technical and experimental set-up details and, finally evaluation of knowledge section \ref{evaluation} which provides methodologies in which knowledge gain is calculated for each of the cognitive learning level. The entire question set for each of the search tasks can be found in Appendix \ref{sec:questionBank}.

\section{Motivation for Proposed Design}\label{sec:motivationDesign}
As it has been discussed previously and will be discussed throughout this thesis, the revised Bloom's taxonomy\citep{krathwohl2002revision} has relaxed constraints between the cognitive processes. This implies that each cognitive process has an overlap in boundaries with its immediate top and bottom cognitive processes in the taxonomic structure. The cognitive processes can be arranged as a hierarchical pyramid as seen in \ref{fig:blooms_taxonomy}, however, due to its relaxed nature of boundaries it can be inferred that if a learner appears for both \textit{Remember} and \textit{Understand}, he is already carrying some existing knowledge from previous level. While this is the motivation in classroom learning scenario, i.e., to have a smooth transition between cognitive processes, it might create imprecise results while measuring user behavior in online learning. \citep{jansen2009using} shows this behavior in \ref{fig:jansen's_search_learning_relation} where it was hypothesized that the higher level tasks did not call for many searches as user already possessed the knowledge from lower levels and hence, required mere verification. Further, \citep{ghosh2018searching} speculates the reason behind the lack of statistically significant difference in user behavior was the ascending order in which tasks were provided to users which made users familiarized with the topic before they reached higher cognitive level.

Therefore, we realized that the study design should be able to distribute tasks among its participants in a special manner such that it is able to overcome the contamination that is added to the results for each category of task due to the continuous nature of cognitive levels of learning in revised Bloom's taxonomy.

\subsection{Research Questions and Hypotheses}\label{sec:RQandHypo}

With the help of this modified design we try to answer three research questions mentioned in section \ref{sec:researchQues}. To answer these research questions, we will use the user data collected from a study designed with the motivation discussed in \ref{sec:motivationDesign} to support the following hypotheses:

To answer \emph{RQ1}, we formulate following hypotheses:

\ul{\textit{Hypothesis 1.1:} Users exhibit changes in knowledge gain for the search tasks of varying cognitive learning levels}

We followed the assumption that if a learner performs a learning task with search sessions, he is bound to gain some knowledge as he proceeds with the task. In order to support \textit{Hypothesis 1.1}, we designed the search tasks and experimental set-up according to proposed framework discussed in following section \ref{sec:design} and recruited workers from a crowd-sourced platform. The search tasks were used to measure their knowledge gain(K.G.) throughout the task session.

\ul{\textit{Hypothesis 1.2:} The change in knowledge gain is dependent upon the cognitive learning level of the search task}

The cognitive learning level of the search task would have some kind of impact on the changes in knowledge gain of the learner.

\ul{\textit{Hypothesis 1.3:} The increase in knowledge gain is dependent upon the hierarchy of cognitive processes of the search task}

As the search tasks of higher complexity learning level will ask the user to solve more complex questions, it can be assumed that this may lead to either a higher knowledge gain when compared with a search task of lower complexity or lower knowledge gain because of the tasks being more complex. Either way, we hypothesize that there should be an upward or downward trend in the increase in knowledge gain when compared across varying cognitive learning levels.

To answer \emph{RQ2} we formulated following hypotheses:

\ul{\textit{Hypothesis 2:} Search behavior in terms of user interactions increases with the increase in cognitive learning complexity of the task}

Since, \citeauthor{krathwohl2002revision}'s taxonomy states that as one climbs higher in the cognitive processes pyramid, the task associated with it becomes more complex. Hence, it is fair to assume that the user interactions of a user attempting a more complex task would be more when compared to that of a user attempting a less complex task as more complex questions would require greater effort to find the answers.
To support \textit{Hypothesis 2}, we need to first prove the following sub-hypothesis.

\ul{\textit{Hypothesis 2.1:} Search queries will increase in number with the increase in cognitive learning complexity of the task}

\ul{\textit{Hypothesis 2.2:} Query length will increase with the increase in cognitive learning complexity of the task}

\ul{\textit{Hypothesis 2.3:} Number of unique query terms will increase in number with the increase in cognitive learning complexity of the task}

\textit{Hypotheses 2.1, 2.2,} and \textit{2.3} focus on query related aspects of the search. As the tasks become more complex, the questions of the tasks see an increase in complexity. Due to this reason, the behavior related to queries like number and length of query should see an increase. 

\ul{\textit{Hypothesis 2.4:} Number of websites visited will increase with the increase in cognitive learning complexity of the task}

\ul{\textit{Hypothesis 2.5:} Number of search pages visited will increase with the increase in cognitive learning complexity of the task}

\textit{Hypotheses 2.4} and \textit{2.5} focus websites and search pages visited. More complex tasks should ideally require a user to refer to many resources before finding the answer for the questions of the task. Hence, there should be an increase in number of websites and search pages visited as a user attempts a more complex task than a lower complexity task.

\ul{\textit{Hypothesis 2.6:} Time spent online will increase with the increase in cognitive learning complexity of the task}

The assumption that led to \textit{Hypothesis 2.6} was that a more complex task would require the user to spend more time online in order to solve it when compared to less complexity task

In order to support \textit{Hypothesis 2} and all it's sub hypotheses, we will use the logged data of user interactions of participants who perform these tasks from a crowd-sourced platform.




\section{Design}\label{sec:design}
The proposed design is to create questions for six different search tasks for each of the six cognitive learning level. Six search tasks for ``Vitamin and Nutrients" domain were created. Each search task corresponded to \citeauthor{krathwohl2002revision}'s cognitive learning processes. These search tasks were hosted on a crowd-sourcing platform called figure-eight\footnote{https://www.figure-eight.com/}. As each of the cognitive level is not independent of each other, we paired a new unique user to only one of the search tasks. This setup gave us the flexibility of providing the users with questions that corresponded to a unique cognitive level. The user was blocked from attempting any further tasks in future. Blocking the user from any future tasks and allowing him to perform for only one task of a specific cognitive learning level ensured that there was no carry over of knowledge from one cognitive learning level to other. This also ensured that for a higher cognitive level, the user will have to first familiarize himself with the topic and carry out research instead of a mere fact-verification as seen in \citep{jansen2009using}. The tasks were added on the platform in a consecutive manner and the users who appeared in precursory tasks were blocked from any and all successive ones irrespective of the fact whether the user carried out a valid or invalid submission. This ensured that any new user who shall appear in succeeding tasks will start with a blank state and have no idea of the domain topic.

\subsection{Experimental Setup}
The six tasks labeled \textit{Remember, Understand, Apply, Analyze, Evaluate,} and \textit{Create} were uploaded on figure-eight which is a crowd-sourcing platform in a consecutive manner. We chose a crowdsourcing platform over laboratory experiments as the study requires a large number of participants\citep{gadiraju2017crowdsourcing}. Further, we were only interested in participants whose native language was English and crowdsourcing the search tasks gave us the flexibility to reach more number of native English speakers.

The title of jobs for all these tasks on the platform was kept uniform - ``Search and Answer", so the users will not have any prior knowledge of the type of work that the search tasks demand. In addition to having a general title, the description of the task too was kept non-specific. A classic description for most of the tasks looked like ``In the task you will answer a few questions and use our custom search engine. The topic for questions will be introduced once you click the task link. You can search for answers when you do not know them using our search engine.
IMPORTANT: The task requires you to have proficiency in English language''. All these cautions were taken so that it will not bring in any bias in results from reading the title and description\citep{hube2019understanding}. Each task had 30 minutes as maximum allocated time. \citep{han2019all} demonstrates that workers on figure-eight often abandoned tasks for the lack of reward, difficulty, and clarity in the task instructions. For this reason, in order to motivate the workers to finish the task, a pay for 50 cents was set for the tasks. Additionally, the workers were given an incentive of a bonus equivalent to 1 US dollar if they performed competently. Further, certain quality control measures were set on figure-eight for all the tasks. A worker was allowed to submit 1 judgment per task, only Level-3\footnote{Level-3 workers on figure-eight are highest Quality workers. It is a group of most experienced, highest accuracy contributors} workers were allowed to attempt the task and we limited the workers to be from English-speaking nations so the workers understand the instructions and questions with full clarity\citep{gadiraju2015understanding, gadiraju2017clarity}.

When a user would click on the search task to attempt it on figure-eight, he will have to click on the task-link which will redirect the user to a different platform in a new tab where the search tasks for the ongoing cognitive learning level were hosted. Here, the user is provided with further instructions on how to attempt the given task along with a small introductory passage describing the importance of ``Vitamin and Nutrients'' in a healthy diet. The instructions informed the user to use \textit{SearchWell\footnote{http://searchwell.l3s.uni-hannover.de/}} search engine exclusively for any search related actions. The user can attempt the task after reading the instructions and upon a valid submission he receives a completion code. In order to get paid on figure-eight they will have to provide this completion code on the platform. Care was taken that the search task platform would not display task to any user who has previously tried to attempt. The validity of submission was determined by the rule that if a user submits a task without carrying out a search and the task contains incorrect answers then the submission is automatically rejected. Also as the aim of our work is to further the understanding of
how the relation between user, search, learning online, it made sense that we discard those users who did not enter a
search query. For \textit{Evaluate} and \textit{Create} level, due to open-ended nature of the tasks, any submission without issuing a single search query was rejected. We will discuss the design of each tasks especially concerning developing the questions as well as the online setup in the subsequent sections. 246 submissions were collected in total by the completion of last search tasks. These submissions included 150 ACCEPTED submissions and 96 REJECTED submissions.

\subsection{Technical Framework and Background}
\citep{Gadiraju:2018:NAS:3266231.3266235} introduces the search environment called \textit{SearchWell}. \textit{SearchWell} is built on top of the Bing Web Search API. It uses tracker\footnote{http://learnweb.l3s.uni-hannover.de/tracker/} to log and track user activities on the
platform including mouse movements, clicks, key presses, URLs visited, time spent on URLs, etc. This recording tool was developed with a WAPS proxy server so the user can continue his online activities without any hindrance from the logging actions. The general design for the tracker is as seen in figure \ref{fig:tracker}. The tracker logs following information:
\begin{figure}
    \centering
    \includegraphics{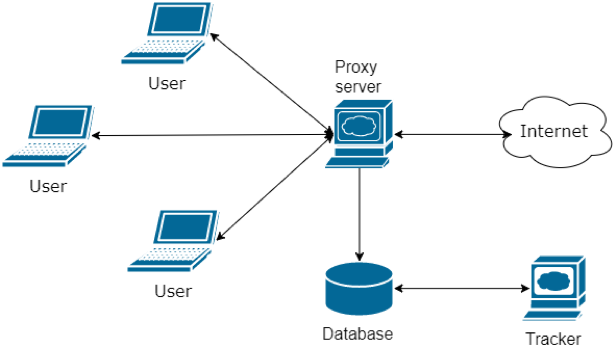}
    \caption{System Design for tracker}
    \label{fig:tracker}
\end{figure}

\begin{itemize}
    \item User's website navigation activities
    \item Time spent on pages, including active and passive(total) times
    \item User's mouse actions like movement of cursor, clicks (button clicks, URL clicks), and position of mouse
    \item Text input in text fields, especially used in recording search queries
    \item Other actions like scrolling, re-sizing windows, key presses, etc
\end{itemize}

The above logs are stored in a MySQL database and used by tracker in recreating user activities.

\begin{figure}
    \centering
    \includegraphics[width=0.75\textwidth]{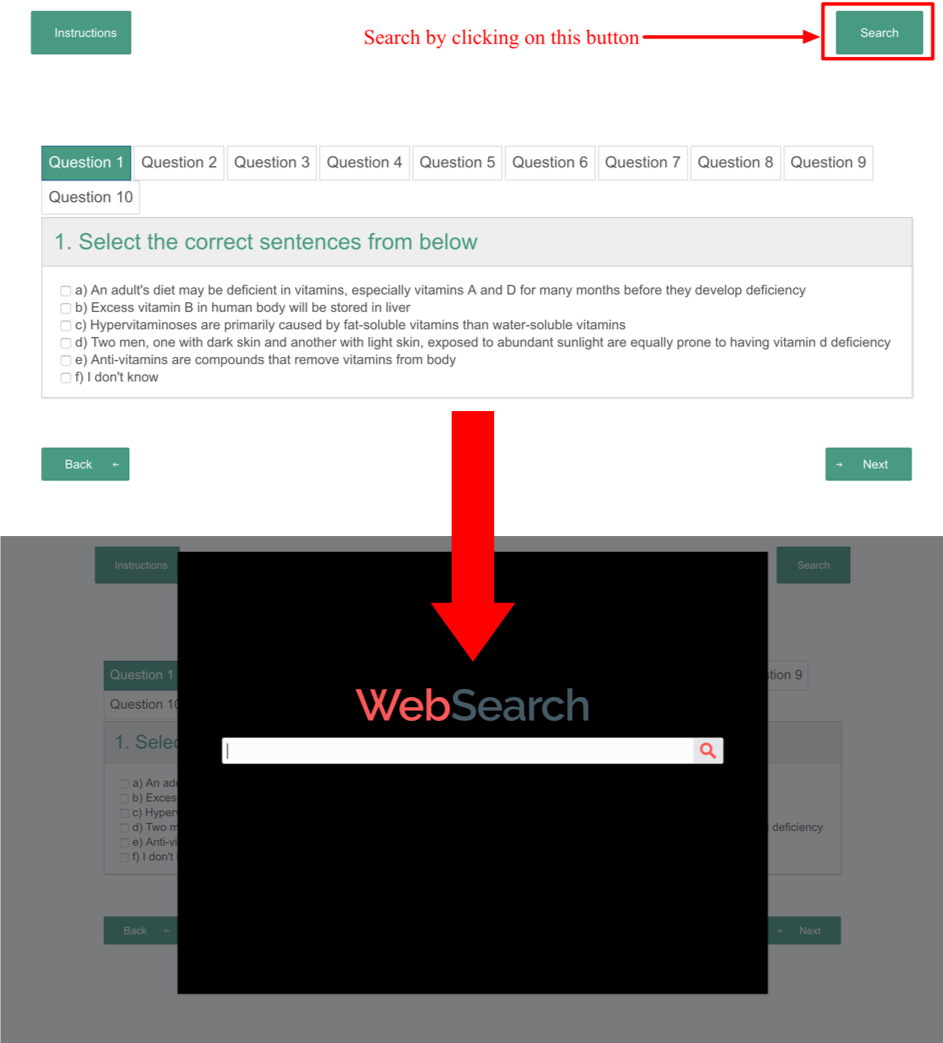}
    \caption{Search interface for the tasks of varying cognitive processes}
    \label{fig:searchInterface}
\end{figure}

In this thesis, we developed a Dynamic Web Project to host the search tasks for each of the cognitive learning levels online. For each of the search tasks, a JSF web-page is developed to host the questions corresponding to the tasks. In addition to this, the web-page for each task, except that of \textit{Remember} includes a button to open \textit{SearchWell} in an iframe. \textit{SearchWell} along with tracker tracks all the user interactions. Figure \ref{fig:searchInterface} shows how the search interface would appear for \textit{Apply} task. This design layout is kept consistent except for \textit{Remember} search tasks but we will discuss more on that in the following section \ref{sec:taskDesign} when we talk about individual task designs.

\subsection{Task Design}\label{sec:taskDesign}
With the help of previous literature \citep{krathwohl2002revision, ferguson2002using, lord2007moving, jansen2009using, kelly2015development} we designed task questions that would require the users to use the action verbs corresponding to each of the cognitive learning level while answering. Verbs or action words reflect the type of action to be carried out on knowledge, for example recall of a fact, providing a judgment, etc. Table \ref{tab:actionWords} provides an overview of mapping of each cognitive level of \citeauthor{krathwohl2002revision}'s taxonomy to the words that can be used to design the questions for the chosen cognitive level. Section \ref{sec:questionBank} provides the entire question bank for all the tasks. It can be seen from this question bank that \textit{Remember} has fifteen questions, \textit{Understand} has ten questions, both \textit{Apply} and \textit{Analyze} has nine questions, \textit{Evaluate} has two questions, and \textit{Create} has only one question. The reason behind the inequality in the number of questions was introduced because as we go higher in the complexity pyramid, the difficulty of the task increases. This would mean that having fifteen questions for \textit{Create} level would be unreasonably difficult, time-consuming, and tedious  in comparison to \textit{Remember}. Hence, in order to bring some kind of equality between the tasks, and to have realistic submission goals from crowd-sourced participants, it did not seem fair to have the same number of questions for all search tasks. Table \ref{tab:taskSetup} gives an overview of the setup for search tasks for all the cognitive learning levels.

\begin{table}[]
    \centering
    \begin{tabular}{L{2cm}L{2cm}L{2cm}L{2cm}L{2cm}L{2cm}}
    \toprule
         \textbf{Cognitive level} & \textbf{Max task length } & \textbf{Pay} & \textbf{Bonus} & \textbf{No. of questions} & \textbf{Max K.G. possible} \\\midrule\midrule
         \textbf{Remember} & 30 min & 0.50 cents & 1.0\$  & 15 & 20 \\\midrule
         \textbf{Understand} & 30 min & 0.50 cents & 1.0\$  & 10 & 24 \\\midrule
         \textbf{Apply} & 30 min & 0.50 cents & 1.0\$  & 9 & 19 \\\midrule
         \textbf{Analyze} & 30 min & 0.50 cents & 1.0\$  & 9 & 47 \\\midrule
         \textbf{Evaluate} & 30 min & 0.50 cents & 1.0\$  & 2 & - \\\midrule
         \textbf{Create} & 30 min & 0.50 cents & 1.0\$  & 2 & - \\
         \bottomrule
    \end{tabular}
    \caption{Task setup for all cognitive learning levels}
    \label{tab:taskSetup}
\end{table}

\begin{table}[]
    \centering
    \begin{tabular}{p{0.2\textwidth}p{0.5\textwidth}p{0.2\textwidth}}
    \toprule
        \textbf{Cognitive level} & \textbf{Verbs/Action words} & \textbf{Potential task}
         \\\midrule\midrule
         \textbf{Remember} & recognize, recall, repeat, state, define, identify, name, list  & Recall information and basic concepts \\\midrule
         \textbf{Understand} & classify, summarize, infer, explain, exemplify, identify, locate, recognize, report, select, describe  & Explain ideas or concepts \\\midrule
         \textbf{Apply} & solve, use, interpret, schedule, execute, implement, demonstrate, operate, sketch & Use the  information in new situations \\\midrule
         \textbf{Analyze} & differentiate, organize, attribute, relate, compare, contrast, distinguish & Draw connections among ideas \\\midrule
         \textbf{Evaluate} & justify, check, critique, weigh, support, judge, defend, argue, appraise & Justify a stand or decision \\\midrule
         \textbf{Create} & create, generate, plan, produce, design, construct, assemble, develop, conjecture, formulate, author & Produce new or original work \\
         \bottomrule
    \end{tabular}
    \caption{Cognitive learning processes mapped to verbs and typical action required }
    \label{tab:actionWords}
\end{table}

\subsubsection{Remember}

For the \textit{Remember} task, 15 questions in the domain ``Vitamins and Nutrients'' were formulated. All these fifteen questions were of either \textit{fill-in-the-blank} type or of \textit{True/False} type. The entire set of questions are listed in the subsection \ref{sec:RememberQuestion} of section \ref{sec:questionBank} which contains the entire question bank. All the questions asked for simple facts as answer which can be found through simple single search queries.

We used the same setup for the \textit{Remember} task as that used in \citep{gadiraju2018analyzing, yu2018predicting}. The task itself was designed to be attempted in three stages. The users were asked to attempt all the three stages in one sitting and weren't allowed to take a break. The time allocated for the task was 30 minutes. The task was hosted on figure-eight for a pay of 50 cents and bonus incentive was provided for an added 1 US dollar.

The user who would like to attempt the search task would click on the task link on figure-eight platform and a new tab for the search task would open. Here, the user is given more information on how to attempt the task as well as an introductory paragraph on the topic on which the questions are based upon which is ``Vitamin and Nutrients''. He is also informed that there are three stages of task namely, 1) Pre-test,
2) Search session, and 3) Post-test. The user is instructed to not carry out any search outside the step 2), i.e., Search session. The questions in pre-test and post-test were kept same in order to measure recall, however, the users were not informed about this fact. They were kept in dark so the search session would be like a learning scenario. In order to get honest results, user is instructed on both, figure-eight instruction area, as well as in the instructions that open up for \textit{Remember} task page that the bonus does not depend upon their existing knowledge but on how honestly they perform. Moreover, they were also informed that if they cheat in the task by searching from anywhere except in the search session, it could lead to rejection. We hoped that this would encourage users to submit more honest answers.
\begin{figure}
    \centering
    \includegraphics{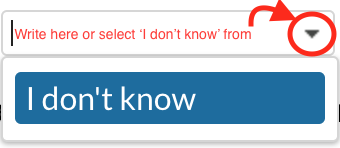}
    \caption{Input options for a text-field of questions in \textit{Remember} search tasks}
    \label{fig:inputRemember}
\end{figure}

\begin{figure}
    \centering
    \includegraphics[width=\textwidth]{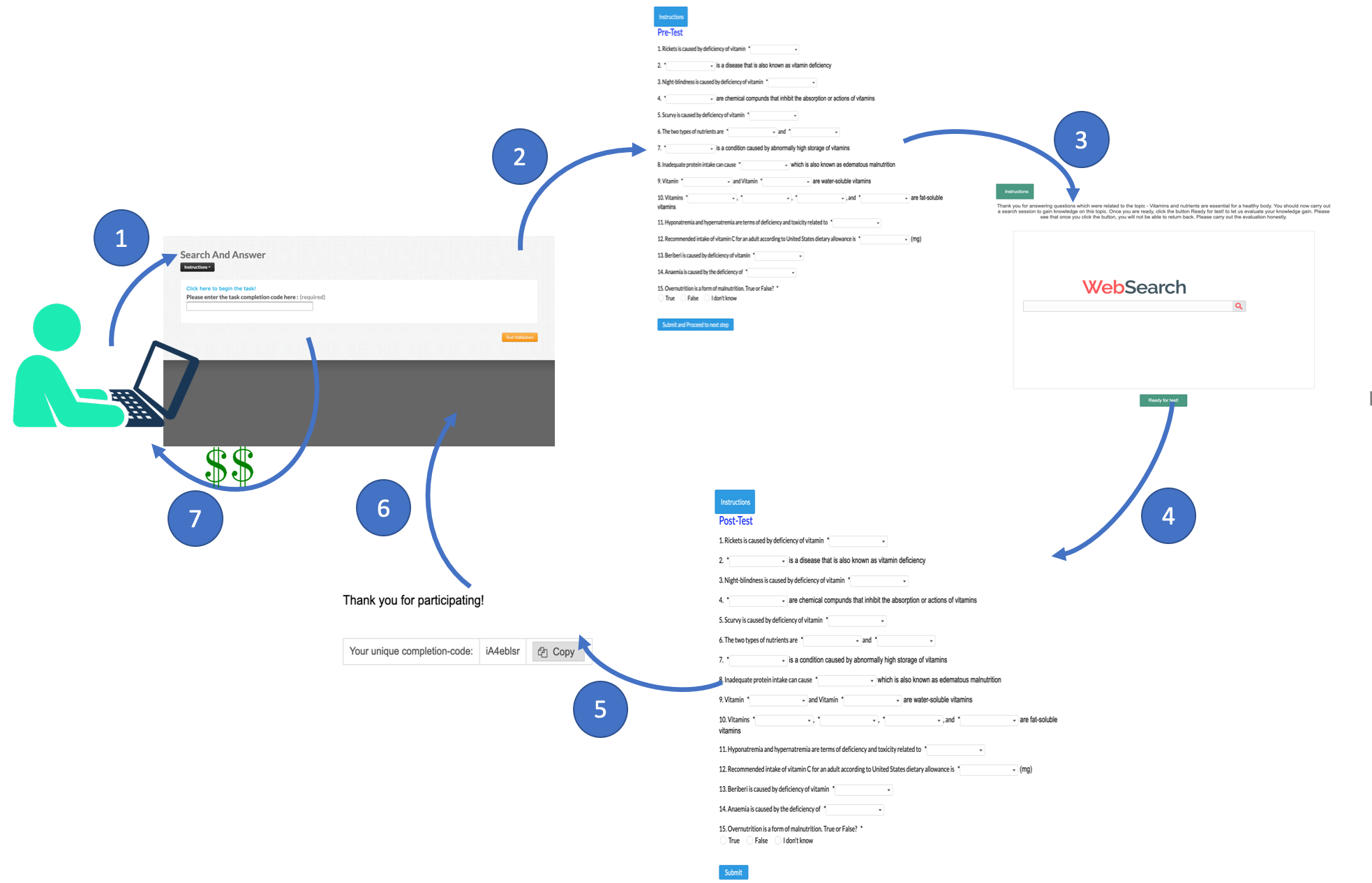}
    \caption[Work-flow for \textit{Remember} search task]{Work-flow for \textit{Remember} search task: 1) Worker is recruited from figure-eight. 2) Worker attempts pre-test by clicking on task-link from figure-eight. 3) Worker carries out search session. 4) Worker feels ready for post-test and proceeds to attempt it. 5) Worker receives a completion code on submit. 6) Worker goes back to figure-eight with completion code. 7) Worker gets paid. }
    \label{fig:rememberWorkflow}
\end{figure}

The answers to questions in pre-test can be either in the form of a simple text input or the user can select ``I don't know" which is displayed upon clicking a text-field. Figure \ref{fig:inputRemember} shows the instruction for input of a typical \textit{Remember} question. Once, the users finished the pre-test, they were able to click a button and continue to second stage which was search session. Here, the users could carry out search from \textit{SearchWell} and gain knowledge on the topic ``Vitamins and Nutrients". The users were instructed that they will not be able to return to this page once they begin post-test and hence, they should spend time in the search session to gather some knowledge on the topic. Once, they felt ready they continued to the final, post-test stage. Here, the users were given the same questions as those in pre-test. Upon submission, any user who submitted without carrying out any search queries and also marked incorrect answers were immediately flagged and rejected. All the other users received a completion code which they were required to copy and paste upon returning back to the figure-eight platform in order to get paid. Figure \ref{fig:rememberWorkflow} gives an overview of the work-flow of the \textit{Remember} task.

We received a total of 56 submissions for \textit{Remember} task from figure-eight. Of these 56, 25 submissions were accepted. 16 submissions were discarded because they did not carry out any search queries in the search session and submitted incorrect answers in post-test. 15 users abandoned the task in between. These were the users who carried out at-least one of the three stages of the task but did not finish it entirely.

\subsubsection{Understand}
For the \textit{Understand} level search task, 10 questions in the domain ``Vitamins and Nutrients'' were formulated. All these were multiple choice questions. The entire set of questions are listed in the subsection \ref{sec:UnderstandQuestion} of the section \ref{sec:questionBank} which contains the entire question bank. It is important to design \textit{Understand} questions with great care as distinguishing them from \textit{Remember} is sometimes confusing and difficult. The questions in this level are slightly more complicated than in \textit{Remember} level. Most of the multiple choice questions required user to recognize item/s from the multiple choices that fulfill the question requirements. This meant that the user will need to identify the correct as well as incorrect items in order to answer accurately. Most of the questions were designed so the user will first have to familiarize himself with the concept stated in the question in order to select the correct answer. One such example is ``Which of the following are the common symptoms associated with avitaminosis C?''. In order to answer this question, the user will ideally carry out a search query for avitaminosis C in order to understand the concept and then find its symptoms.

\begin{figure}
    \centering
    \includegraphics[width=\textwidth]{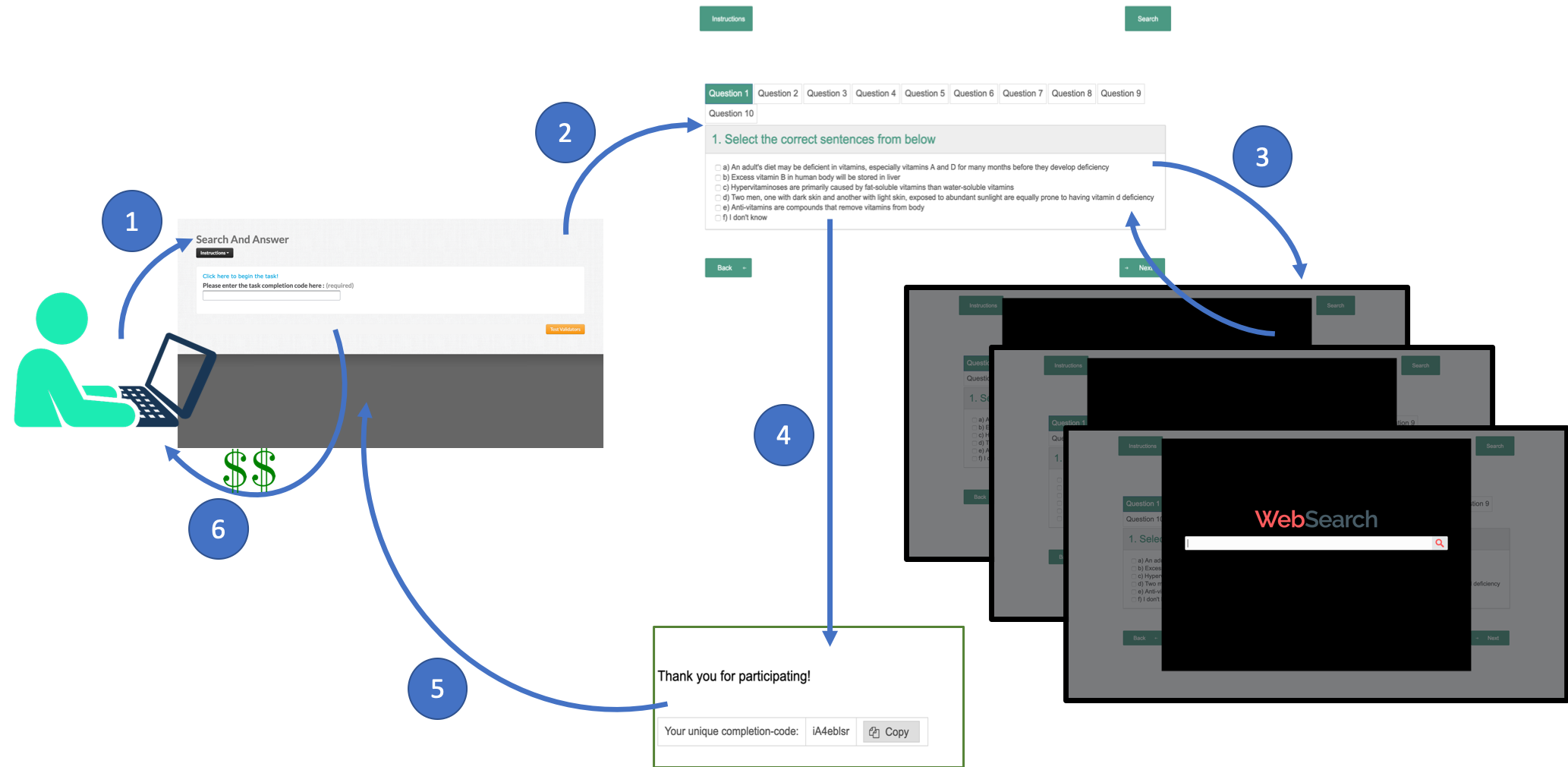}
    \caption[Work-flow for \textit{Understand, Apply,} and \textit{Analyze} search tasks]{Work-flow for \textit{Understand, Apply,} and \textit{Analyze} search tasks: 1) Worker is recruited from figure-eight. 2) Worker attempts the task by clicking on task-link from figure-eight. 3) Worker opens search frame many times as he proceeds with the task and encounters new questions.  4) Worker receives a completion code on submit. 5) Worker goes back to figure-eight with completion code. 6) Worker gets paid. }
    \label{fig:understandWorkflow}
\end{figure}

\textit{Understand} task was published on figure-eight after \textit{Remember} got completed. All those users, whether with valid or invalid entries, who submitted for \textit{Remember} were blocked from attempting this task with a simple JavaScript as we require unique workers for each of the task. The user who would like to attempt the search task would click on the task link on figure-eight platform and a new tab for the \textit{Understand} search task would open. The entire set-up of instructions and pay is similar to \textit{Remember}. The user is motivated to execute the search task honestly through instructions, both on figure-eight and on task page where it is mentioned that assignment of bonus for the task does not depend on their existing knowledge and that they are encouraged to search and not guess the answers. For each of the question, the user needs to give an answer mandatory. He can however, click "I don't know" as it is provided as one of the choices for the answer. The user can carry out search at any time during the period in which he is performing the task. Ideally, the user would carry out search to find answers for each of the question. The search frame would open \textit{SearchWell} in an iframe where user will try to find correct answers. Upon finding the answer, he would return to the task web page and select the correct options. Figure \ref{fig:understandWorkflow} shows the work-flow of a user while attempting \textit{Understand task}.

We received a total of 40 submissions by users from figure-eight. Of these 40 submissions, 25 were accepted while 15 were rejected as these were the users who submitted answers without carrying out any search queries as well as incorrect answers. 

\subsubsection{Apply}
For the \textit{Apply} level search task, 9 questions in the domain ``Vitamins and Nutrients'' were formulated. The entire set of questions are listed in the subsection \ref{sec:ApplyQuestion} of the section \ref{sec:questionBank} which contains the entire question bank. Questions in \textit{Apply} are either of ordering type or multiple-choice type questions. Ordering type questions describe a scenario and ask the user to order the sentences that lead to the said scenario. MCQs for \textit{Apply} are formulated differently then those of \textit{Understand}. The questions of MCQs of this level require user to understand the concept and then apply the knowledge to realize the correct answer. For example, a question like ``Mary frequently gets muscle pain especially in her legs during night. Mary's symptoms are most likely associated with which vitamin deficiency?" would first require the user to find vitamin deficiencies of various vitamins and understand the symptoms and then apply in Mary's case. It is crucial to design the questions for each level where the action of cognitive level is used while answering.

The work-flow and task setup is same as that of \textit{Understand} level and as shown in figure \ref{fig:understandWorkflow}. All the users, whether with valid or invalid entries, who submitted previously were blocked from attempting this task as we require unique workers for each of the task. The users are encouraged to search whenever they are unsure of answers instead of guessing. Any user who submitted without carrying out a single search query during the entire task session and had incorrect answers in the submitted result was flagged and his work was rejected from being considered for this thesis. We received a total of 35 participants of which 25 had valid submissions and work of 10 participants was rejected.

\subsubsection{Analyze}
\begin{figure}
    \centering
    \includegraphics[width=\textwidth]{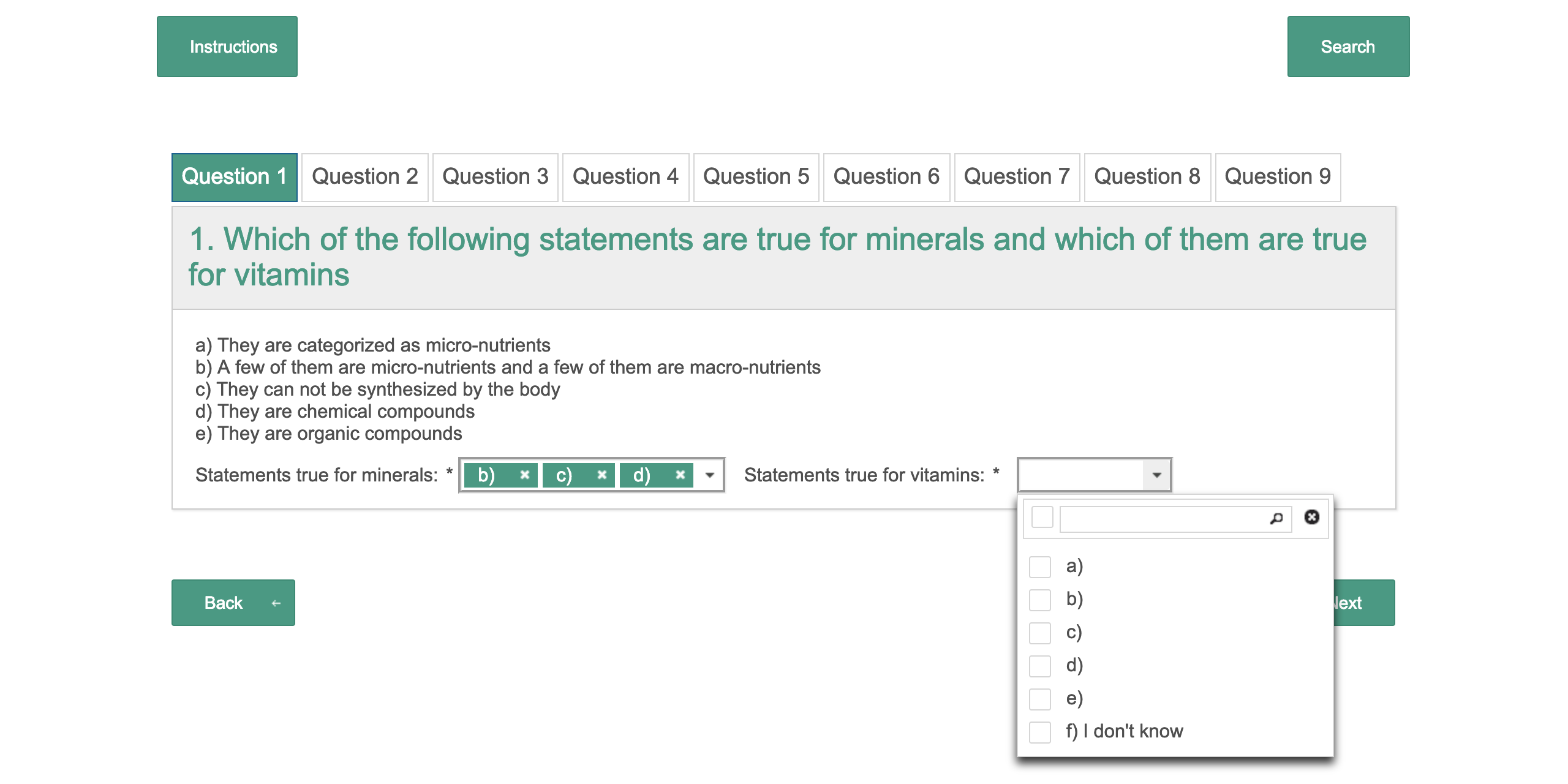}
    \caption{Typical analyze question and how it looks on the task link page}
    \label{fig:analyzeQuestion}
\end{figure}

For the \textit{Analyze} level search task, 9 questions in the domain ``Vitamins and Nutrients'' were formulated. The entire set of questions are listed in the subsection \ref{sec:AnalyzeQuestion} of section \ref{sec:questionBank} containing the entire question bank. Most of the questions of \textit{analyze} level used action words like differentiate, compare, and contrast. A typical example of an \textit{analyze} question is ``Which of the following statements are true for minerals and which of them are true for vitamins'' followed by a list of statements. The user then attributes each statement to fall either under mineral category or under vitamin category. Figure \ref{fig:analyzeQuestion} shows how a figure-eight user would see a typical analyze question on the task link page.

The work-flow and task setup is same as that of \textit{Understand} and \textit{Apply} level. Figure \ref{fig:understandWorkflow} represents the work-flow of worker from figure-eight to \textit{Analyze} task link page and back to figure-eight to receive the payout. All the users, whether with valid or invalid entries, who submitted for any of the previous tasks were blocked and new users are encouraged to search whenever they are unsure of answers instead of guessing. Any user who submitted without carrying out a single search query during the entire task session and had incorrect answers in the submitted result was flagged and his work was rejected from being considered for this thesis. We received a total of 47 submissions of which 25 had valid submissions and work of 22 participants was rejected.

\subsubsection{Evaluate}
\textit{Evaluate} task consisted of two sub-questions for a given question. The task described a scenario and asked the user to provide a judgment. In addition, the user was asked to support his judgment by carrying out research and providing a scientifically sound answer. This task is an open-ended task where user is asked to explore the web and gather information in order to back up his reasoning. A detailed question in the domain ``Vitamins and Nutrients'' was formulated for this level which will call for a user to use his evaluation skills. The question along with sub-questions are listed in the subsection \ref{sec:EvaluateQuestion} of the section \ref{sec:questionBank}. While formulating the questions for \textit{evaluate} task, it is important to design a question which does not have an answer readily available on the web. For example, ``Is absence of vitamins from diet good or bad for you?'' albeit being a judgmental type question does not form a good \textit{Evaluate} task question as a simple search query would reveal a Wikipedia page providing the answer for the question.

\begin{figure}
    \centering
    \includegraphics[width=\textwidth]{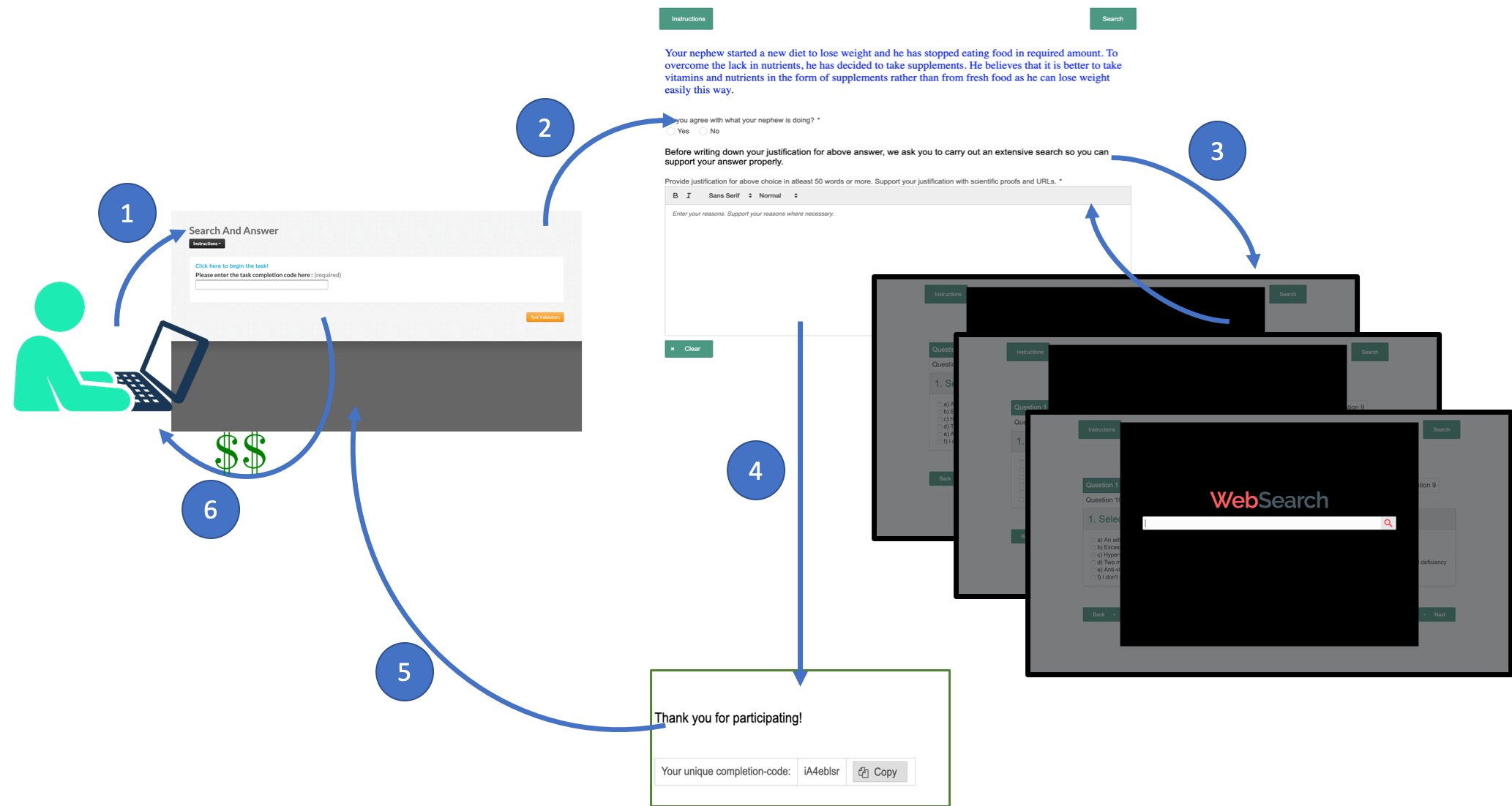}
    \caption[Work-flow for \textit{Evaluate} and \textit{Create} search tasks]{Work-flow for \textit{Evaluate} and \textit{Create} search tasks: 1) Worker is recruited from figure-eight. 2) Worker attempts the task by clicking on task-link from figure-eight. 3) Worker opens search frame many times while answering main question/s of task.  4) Worker receives a completion code on submit. 5) Worker goes back to figure-eight with completion code. 6) Worker gets paid. }
    \label{fig:evaluateWorkflow}
\end{figure}

The task setup on figure-eight is same as that of \textit{Remember, Understand, Apply,} and \textit{Analyze} level. All users with previous submissions were blocked. The users for \textit{Evaluate} task were encouraged to carry out extensive search to support their evaluation. Any user who submitted without carrying out a single search query during the entire task session was flagged and his work was rejected from being considered for this thesis. Due to the exploratory nature of the task, any answer without a backing of research on web was rejected.

The work-flow of \textit{evaluate} search task is slightly different as compared to others. Figure \ref{fig:evaluateWorkflow} describes the work-flow of of a contributor from the moment he clicks to attempt the \textit{evaluate} task till he receives a payout. Upon clicking the task link, the user opens the \textit{evaluate} search task page in a new tab where he will read instructions on how to attempt the task as well as an introductory paragraph on the topic ``Vitamin and Nutrients''. Then, the user will proceed to attempt the one descriptive type question that is for the aforementioned task. The user will refer to web many times during the search task session. The user will refer to web for various reasons such as to determine the answer, modify the answer, and support the answer for the asked question. Upon submitting and successfully completing the task, he will receive a completion code which he can paste on the figure-eight platform to receive the payment. Validity checks are added on submission to verify if the user entered the supporting judgment meeting certain basic criteria. We received a total of 31 submissions of which 25 had valid submissions and work of 6 participants was rejected.

\subsubsection{Create}
\textit{Create} task calls for users to create, design, or plan something new. This search task consisted of one question where the user was asked to design a food plan for a specific case. The specific case was provided so the food chart to be designed is for a unique individual with unique needs. This ensured that the answer for the task was not readily available on web. This task, like \textit{Evaluate} search task is an open-ended one. In order to complete the task, the user needs to use all the lower cognitive learning levels as this is the most complex task. The detailed question in the domain ``Vitamins and Nutrients'' that was formulated for this level is listed in the subsection \ref{sec:CreateQuestion}.

\begin{figure}
    \centering
    \includegraphics[width=\textwidth]{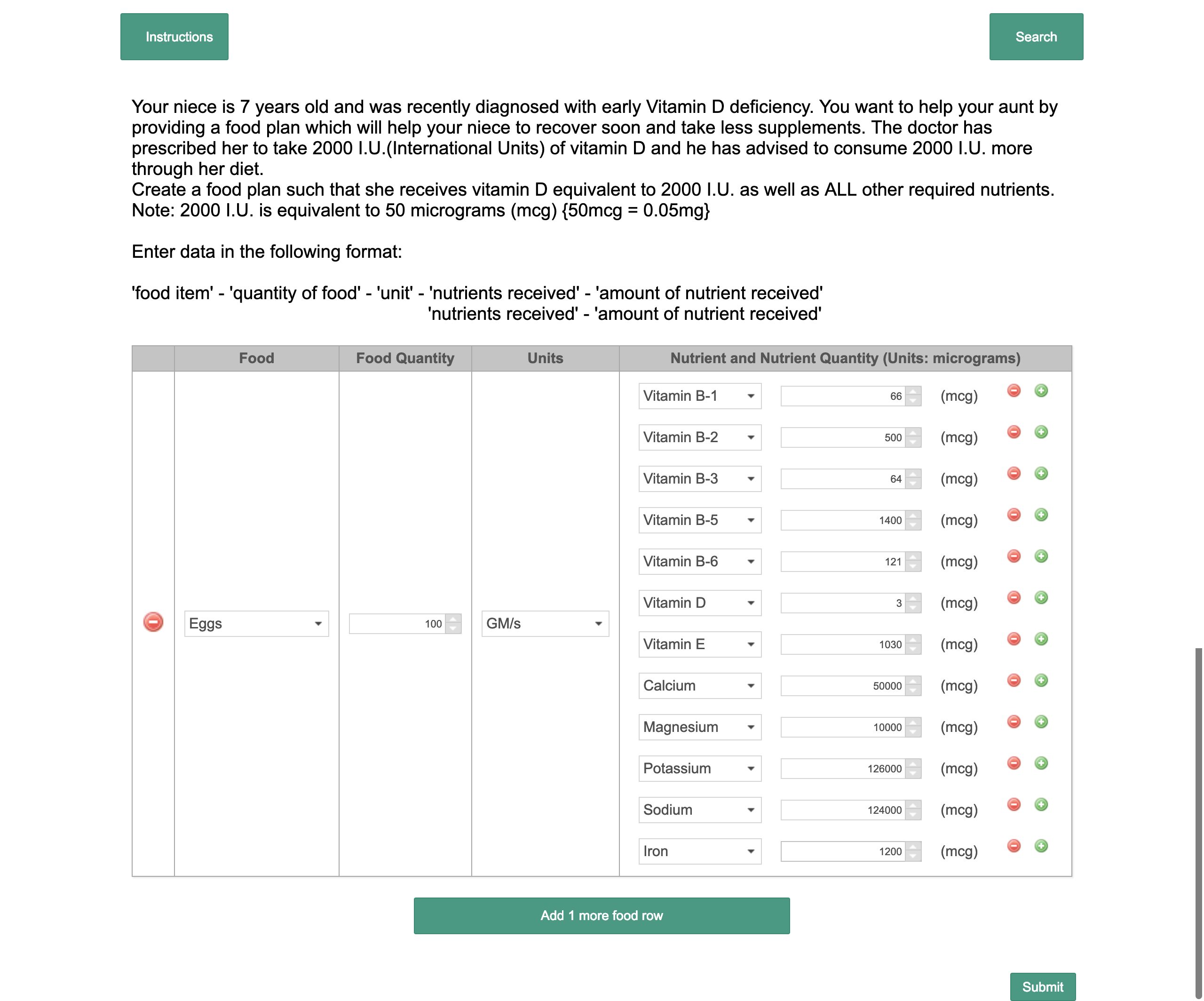}
    \caption{Example of an entry in food plan for 100 GMs of eggs}
    \label{fig:create}
\end{figure}

The task setup on figure-eight is same as that of all the other level. The work-flow of \textit{Create} task is same as \textit{Evaluate} and can be described in figure \ref{fig:evaluateWorkflow}. Any user who tried to attempt previous tasks were blocked from attempting this task as we require unique workers who never attempted the search tasks before. The users are encouraged to search instead of guessing. Any user who submitted without carrying out a single search query during the entire task session was flagged and his work was rejected from being considered for this thesis. Further, validity checks are added on submit, preventing users from continuing to submit a food-plan without the required nutritional values. These validity checks further increases the task complexity. The users designed the food plan by adding various entries that consisted of the food item, quantity of food item intake, nutrients received, and the amount of nutrients received. Figure \ref{fig:create} shows how a user would add an entry of 100 GMs of eggs to the food plan.

Due to the exploratory nature of the task, any answer without a backing of researching on web was rejected. We received a total of 33 submissions of which 25 had valid submissions and work of 8 participants was rejected.

\section{Evaluation of Knowledge}\label{evaluation}
In order to measure learning for each cognitive level, an evaluating metric is designed to calculate the knowledge gain. Table \ref{tab: MeasuringTechniques} provides an overview of the evaluation method for all the cognitive learning levels of \citeauthor{krathwohl2002revision}'s taxonomy. Following sub-sections describe the evaluating procedure for each tasks in detail. While calculating the knowledge gain, negative knowledge is not recognized. Knowledge is either gained or considered to have stayed constant. We do not consider knowledge to have decreased from the the search sessions.

\begin{table}[]
    \centering
    \begin{tabular}{L{2.5cm}L{5cm}L{6cm}}
    \toprule
        \textbf{Cognitive level} & \textbf{Typical task Description} & \textbf{Evaluation} \\ [0.5ex] 
     \hline\hline
      \textbf{Remember}   &  Tasks include Simple questions with straight-forward answers in pre-test and post-test & 1. Incorrect pre-test answer and correct post-test answer for a given question $\implies$ +1\newline 2. Unknown/Unanswered pre-test answer and a correct post-test answer for a given question $\implies$ +1\newline 3. All other cases $\implies$ +0\\
      \midrule
       \textbf{Understand}  & Tasks include Multiple choice questions \newline & 1. +1 for each item correctly selected in the answer\newline 2. -1 for each item incorrectly selected
       \newline 3. Total score =\{correct $-$ incorrect\}\newline =\{0\} if incorrect items $\geq$ correct items \\
      \midrule
        \textbf{Apply} & Tasks include MCQs or re-ordering statements where information in new scenario is applied. & \underline{\emph{MCQs:}}
        \newline 1. +1 for each item correctly selected in the answer\newline 2. -1 for each item incorrectly selected
       \newline 3.  Total score =\{correct $-$ incorrect\}\newline =\{0\} if incorrect items $\geq$ correct items
       \newline \underline{\emph{Ordering:}}
       \newline 1. +1 for correct order
       \newline 2. All other cases $\implies$ 0\\
        \midrule
       \textbf{Analyze} & Tasks include differentiating, comparing, attributing type of questions & \underline{\emph{For each attribute:}}
       \newline1. +1 for each item correctly attributed in the answer\newline 2. -1 for each item incorrectly attributed
       \newline 3.  Total score =\{correct $-$ incorrect\}\newline =\{0\} if incorrect items $\geq$ correct items \\
        \midrule
        \textbf{Evaluate} & Tasks include providing a judgment as well justifying it. & Due to open-ended nature of task, tasks are evaluated manually\\\midrule
        \textbf{Create} & Designing or Creating a plan & Due to open-ended nature of task, tasks are evaluated manually\\
        \bottomrule
    \end{tabular}
    \caption{Knowledge evaluating techniques for each cognitive level}
    \label{tab: MeasuringTechniques}
\end{table}

\subsection{Remember}

The \textit{Remember} task consists of three stages: 1) pre-test, 2) search session, and 3) post-test where pre-test and post-test consists of same questions. Answers submitted in pre-test can be be viewed as existing knowledge of user. The second stage, i.e., the search session is where the user will carry out search queries and gain knowledge. Finally, we will evaluate the knowledge gain for recall cognitive level of the user by comparing the answers submitted in post-test to pre-test.
\begin{itemize}
    \item If a user marked an answer as unknown in pre-test and gave correct answer in post-test, it implies an increase in knowledge
    \item If a user marked an answer as incorrectly in pre-test and gave correct answer in post-test, it implies an increase in knowledge
    \item If a user marked an answer as unknown in both the tests, it implies no increase in knowledge
    \item If a user marked an answer incorrectly in both the tests, it implies no increase in knowledge
\end{itemize}

In order to calculate the increase in knowledge gain, for every instance where an increase was perceived, +1 was awarded.

\subsection{Understand}
\textit{Understand} tasks had questions with multiple correct answers. An answer for a question was considered to be an existing knowledge if the user did not open the search frame while attempting to answer. As the user did not carry any search before answering, it meant that the user is answering the question with the help of his existing knowledge. If the search log indicated that user carried out activities on web while he was attempting the question, then the answers for the question are used to calculate the knowledge gain. The knowledge gained for a question is calculated according to following rules:
\begin{itemize}
    \item For every correctly chosen option of answers for the question, the knowledge gain increases by +1
    \item For every incorrectly chosen option of answers for the question, the total knowledge gain is marked down by -1
    \item Total knowledge gain becomes zero for a question if there are more or equally incorrectly marked options of answers in comparison to correctly marked options of answers
\end{itemize}

\subsection{Apply}
\textit{Apply} task consists of two types of questions - questions with multiple correct answers and questions with statements to be ordered in correct sequence. For both these type of questions, an answer for a question was considered to be an existing knowledge if the user did not open the search frame while attempting to answer. If the search log indicated that user carried out activities on web while he was attempting the question, then the knowledge that is gained is calculated. Rules for questions with multiple correct answers is same as discussed above for \textit{Understand}. However, rules for ordering or sequencing type questions are as follows:
\begin{itemize}
    \item Knowledge is gained by +1 for every correctly answered sequence
    \item Total knowledge remains unchanged for any other scenario
\end{itemize}

\subsection{Analyze}

\textit{Analyze} task questions ask users to attribute different properties to different attributes. Again, like other levels discussed above (except \textit{Remember}), an answer for a question was considered to be an existing knowledge if the user did not open the search frame while attempting to answer. If the search log indicated that user carried out activities on web while he was attempting the question, then the knowledge gain is calculated. The rules for calculating the knowledge gain for a question are as follows:
\begin{itemize}
    \item Knowledge gain is calculated for each attribute
    \item For every correctly assigned property to an attribute, knowledge is gained by +1
    \item For every incorrectly assigned property to an attribute, total knowledge is reduced by -1
    \item Total knowledge gain becomes zero for a question if there are more or equally incorrectly marked options of answers in comparison to correctly marked options of answers
\end{itemize}

\subsection{Evaluate and Create}

Both, \textit{Evaluate} and \textit{Create} have creative, exploratory questions which brings in open-ended answers. As each answer is dependent upon user's thinking and hence, can not be marked as correct or incorrect through a set of rules, we did not calculate the numerical value of increase in knowledge. We carried out manual checking and marked submissions as valid upon encountering complete and comprehensive submissions which made sense and had search logs indicating search actions. We also assumed that users with such submissions would have experienced some kind of knowledge gain.


\chapter{Results and Discussion} 

\label{Chapter5} 

\lhead{Chapter 5. \emph{Results and Discussion}} 


We introduced research questions in Section \ref{sec:researchQues} and formulated hypotheses that would be used to find answers of the aforementioned RQs in section \ref{sec:RQandHypo}. In this section, we will analyze the data collected from the study and discover if the hypotheses are supported with the empirical proof. We performed one-way across subjects ANOVA where task complexity was kept as independent variable to determine if the hypotheses are justifiable.

\section{Relation Between Knowledge Gain and Cognitive Levels}\label{KGRQ1Results}

Research question 1 was aimed at studying changes in knowledge gain online across varying cognitive learning levels. To find the relation between increase in knowledge and the cognitive complexity of the search task, we asked the users, questions based on topic ``Vitamin and Nutrients'' for search tasks of varying cognitive complexity level. We measured the knowledge gained among users based on the answers that they submitted for the task questions. To support the relation between KG and cognitive learning levels, we intend to prove following hypotheses as true:\newline\ul{\textit{Hypothesis 1.1:} Users exhibit measurable changes in knowledge gain for search tasks of varying cognitive learning levels}

\begin{figure}
\centering
\begin{minipage}{0.5\textwidth}
    \includegraphics[width=0.9\textwidth]{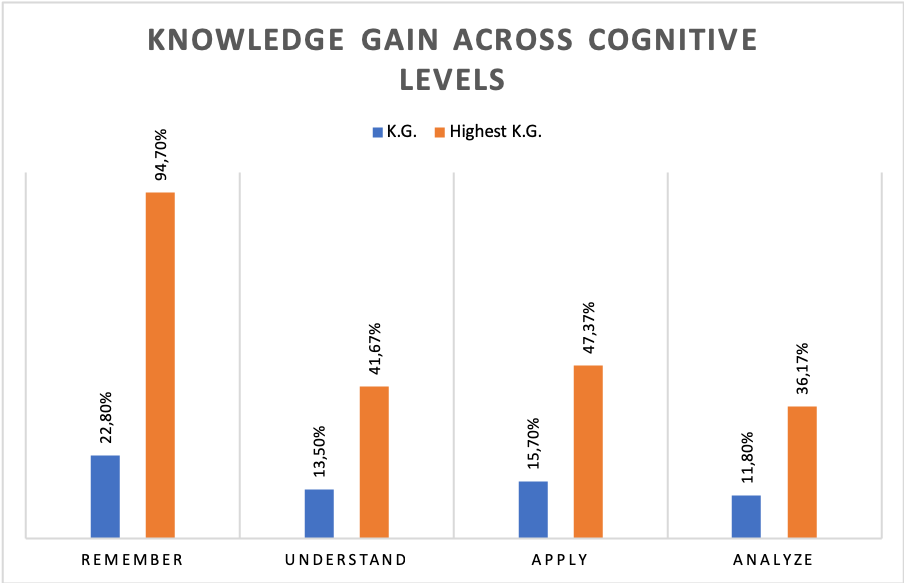}
    \caption{Mean Knowledge gain and maximum knowledge gain across cognitive levels}
    \label{fig:KGCognitveLevel}
    \end{minipage}%
\begin{minipage}{0.5\textwidth}
    \includegraphics[width=0.9\textwidth]{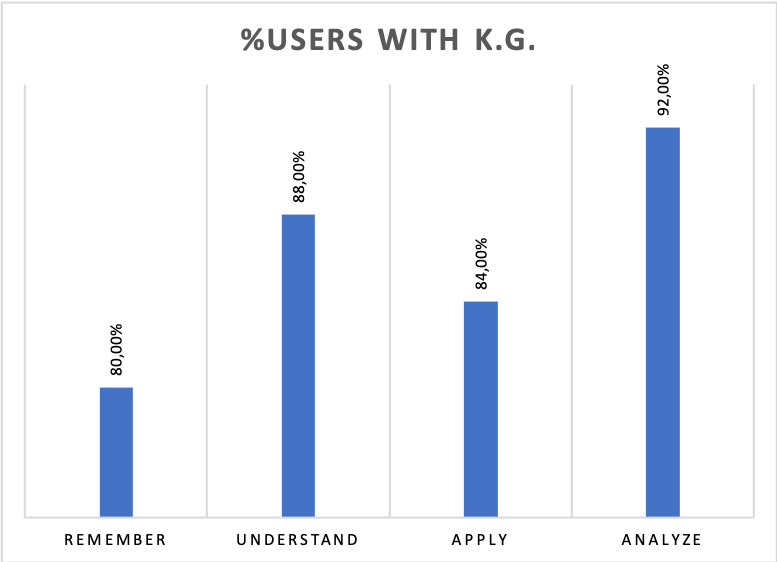}
    \caption{Percentage of users exhibiting knowledge gain for each level}
    \label{fig:usersKG}
    \end{minipage}
\end{figure}

In order to prove Hypothesis 1.1, we carried out crowd-sourced experiments across cognitive level domain of \citeauthor{krathwohl2002revision}. Of the 150 valid, accepted submissions, we calculated the numeric value of knowledge gain (K.G.) for 100 workers across four domains \textit{Remember, Understand, Apply,} and \textit{Analyze}. Of these 100 workers, 86 workers exhibited an increase in knowledge. A minimum knowledge gain of 6\% was shown whereas, a maximum knowledge gain of 94.7\%. Figure \ref{fig:KGCognitveLevel} shows average knowledge gained across the first four cognitive levels as well as maximum knowledge gained across these cognitive levels. Figure \ref{fig:usersKG} shows the percentage of users who exhibited a gain in knowledge for the four cognitive levels. It shows that the majority of the users(\textgreater 86\%) who participated in online learning tasks  experienced knowledge gain which proves learning to be an outcome for each of the chosen cognitive level. Further a user would experience an average knowledge gain between 11\% to 22\% depending upon the cognitive learning level. These statistics prove that knowledge is evolved across the four cognitive domains namely, \textit{Remember, Understand, Apply,} and \textit{Analyze}. 

Due to the open-ended nature of \textit{Evaluate} and \textit{Create} tasks, it is not possible to measure the knowledge gain, however, we believe that the user behavior in terms of gain in knowledge can be extended to the highest two levels as well. We believe so because of the fact that the users of \textit{Evaluate} and \textit{Create} carry out search and spent significant amount of time on web while solving task questions. In addition, looking at search interactions in section \ref{RQ2Results}, we can say that the search behavior for the \textit{Evaluate} and \textit{Create} tasks were comparable to the remaining four domains, and in many cases as shown in following section more than the the lower four domains, due to this, it would have definitely led to an increase in knowledge much like \textit{Remember, Understand, Apply,} and \textit{Analyze}. Further, the manual assessment of answers showed that users submitted valid answers. Therefore, we believe that it is safe to assume a gain in knowledge occurred for \textit{Evaluate} and \textit{Create} tasks for users who searched online and submitted valid entries. Hence, we prove hypothesis 1.1 with empirical proof for first four cognitive learning levels and believe that it holds true for final two cognitive learning levels.

\ul{\textit{Hypothesis 1.2:} The change in knowledge gain is dependent upon the cognitive learning level of the search task}

In order to determine if the cognitive learning level had any impact upon the changes in knowledge gain that is observed for each level, we carried out a one-way between subjects ANOVA for the calculated knowledge gain of first four cognitive level. We calculated the knowledge gain as discussed in Section \ref{evaluation}. We normalize the knowledge gain by the maximum knowledge gain that is possible for each task. This provides us with the percentage of knowledge that is gained. Further, we normalized the user data by the number of questions in each search task to find the knowledge gain per question for tasks of each cognitive complexity. A one-way between subjects ANOVA shows that the increase in knowledge gain per question is affected by the cognitive complexity of the task[$F(4, 150)=21.88, p<0.001$] for normalized data. Results of one-way between subjects ANOVA prove that the increase in knowledge gain is impacted by the cognitive learning level of search tasks. Hence, we show that hypothesis 1.2 is supported with empirical proof for first four cognitive learning level.

\ul{\textit{Hypothesis 1.3:} The increase in knowledge gain is dependent upon the hierarchy of cognitive processes of the search task}

\begin{figure}
\centering
\begin{subfigure}{0.5\textwidth}
  \centering
  \includegraphics[width=0.9\linewidth]{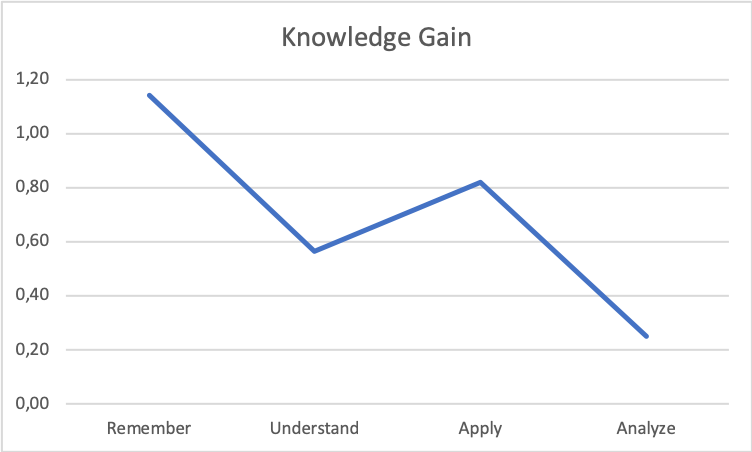}
  \caption{Knowledge Gain\% for cognitive level}
  \label{fig:TotalKGALLQues}
\end{subfigure}%
\begin{subfigure}{0.5\textwidth}
  \centering
  \includegraphics[width=0.9\linewidth]{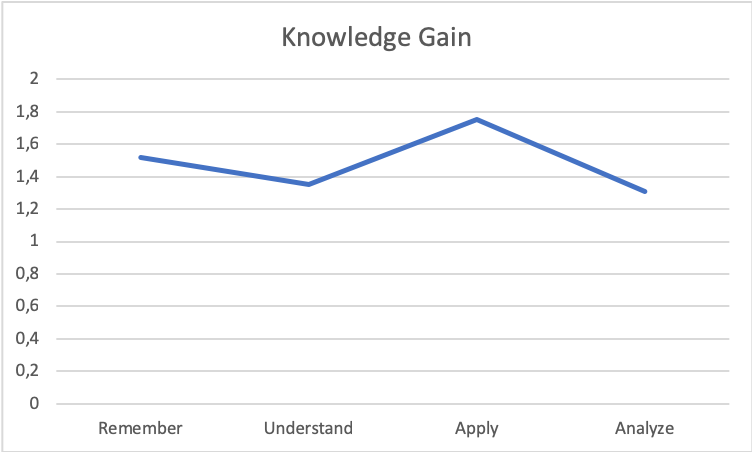}
  \caption{Knowledge Gain\% for user data normalized by number of questions}
  \label{fig:TotalKGPerQues}
\end{subfigure}
\caption{Knowledge Gain Across Cognitive Levels}
\label{fig:TotalKG}
\end{figure}

To support Hypothesis 1.3, we analyzed the relation between the increase in knowledge gain and the cognitive dimension of revised Bloom's taxonomy. We tried to find a relation between KG and the hierarchy of cognitive learning levels. The answers submitted by users for search tasks were used to resolve the hypothesis. We calculated and normalized the knowledge gain by the maximum knowledge gain that is possible for each task to determine the percentage of knowledge that is gained. Figure \ref{fig:TotalKGALLQues} provides an overview of the data.  While there is no significant pattern in figure \ref{fig:TotalKG} between increase in knowledge gain and task complexity, figure \ref{fig:TotalKGALLQues} shows that knowledge gain for \textit{Analyze} a higher complexity level task is significantly lower than \textit{Remember} which is a low complexity task. \ref{fig:TotalKGPerQues} shows a similar but slightly less pronounced trend in the knowledge gained per question for every task. However, these results are insufficient to support hypothesis 1.3.

\section{Relation Between Search Behavior and Cognitive Level of task}\label{RQ2Results}

Research question 2 was aimed at studying the relations between a user's search behavior and the cognitive learning level of of the search task and whether there is any impact on user's interactions by the cognitive learning level of the search task. We hypothesized that:\newline\ul{\textit{Hypothesis 2:} Search behavior in terms of user interactions increases with the increase in cognitive learning complexity of the task}\newline To find a solution for this research question, we collected data related to user's interactions while they performed on search tasks of varying cognitive complexities. The average results are discussed on two factors: 
\begin{enumerate}
    \item Average calculated by considering number of users for each feature
    \item Average calculated by considering number of users normalized by the number of questions in entire search task for each feature
\end{enumerate}

These results are used to support the sub-hypotheses of hypothesis 2 and therefore, also support hypothesis 2.

\subsection{User Queries}

\ul{\textit{Hypothesis 2.1:} Search queries will increase in number with the increase in cognitive learning complexity of the task}

\begin{figure}
\centering
\begin{minipage}{0.5\textwidth}
    \includegraphics[width=0.9\textwidth]{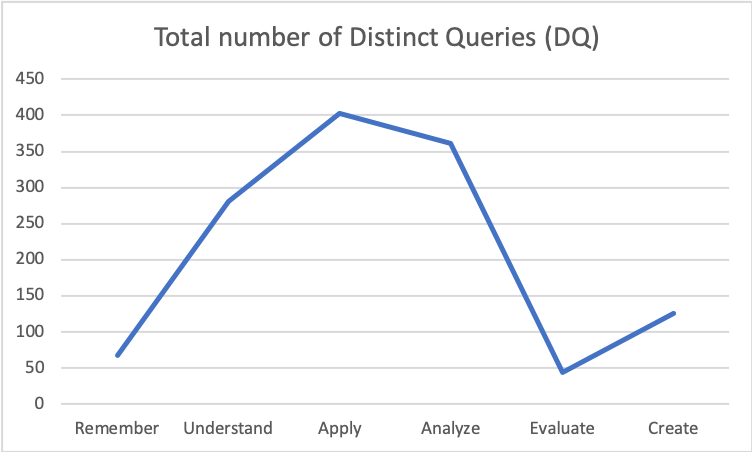}
    \caption{Total number of distinct queries for different cognitive levels}
    \label{fig:TotalDQ}
    \end{minipage}%
\begin{minipage}{0.5\textwidth}
    \includegraphics[width=0.9\textwidth]{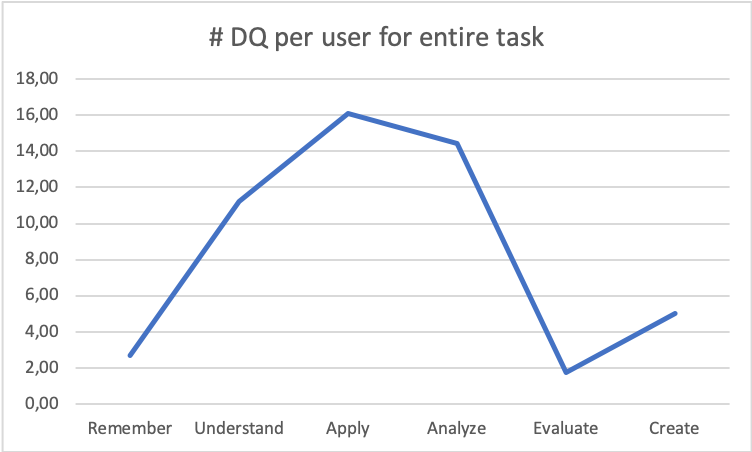}
    \caption{Number of distinct queries per user for every cognitive level}
    \label{fig:usersDQ}
    \end{minipage}
\end{figure}

We collected a total of 1285 distinct queries across all the cognitive levels. The number of distinct queries varied between users from 1 query per user to 36 queries per user. Figures \ref{fig:TotalDQ} shows the total number of distinct queries for each cognitive learning level. Figure \ref{fig:usersDQ} shows number of distinct queries fired per user at every cognitive learning level. Both the curves are similar since the number of participants with valid submissions for each search task is same, which is 25 participants.

\begin{figure}
\centering
\includegraphics[width=0.5\textwidth]{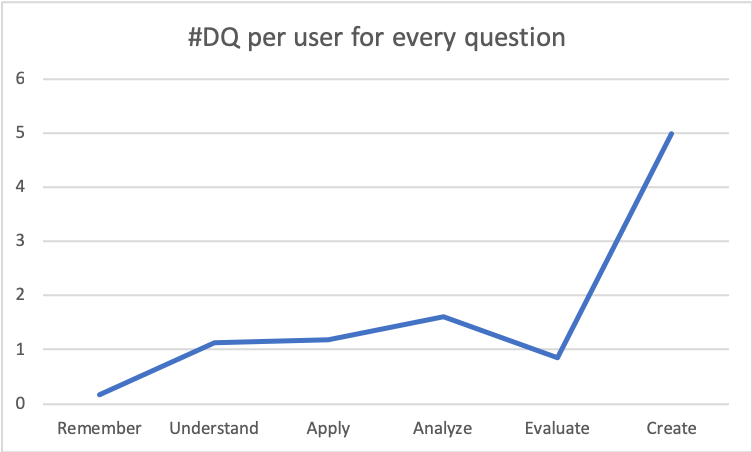}
\caption{Number of distinct queries for every user normalized by number of questions in search task}
\label{fig:userDQquestion}
\end{figure}

Upon normalizing the data by the number of questions in each search task, we gathered the number of distinct search queries per user normalized by number of questions in each task. A one-way between groups ANOVA showed that there were significant statistical differences[$F(6, 150) = 27.68, p<0.001$] for number of distinct queries by each user normalized by number of questions in each search task of varying cognitive learning level. Figure \ref{fig:userDQquestion} shows the trend of increase in distinct queries per question, per user as the complexity of the task increases. The number of distinct queries is lowest for \textit{Remember} and highest for \textit{Create}. However, \textit{Evaluate} creates an anomaly by having number of distinct queries less than \textit{Understand, Apply, }and \textit{Analyze} and falling out of the upwards trend but all the other tasks show an upward growth trend. We believe that \textit{Evaluate} creates an anomaly due to the judgmental nature of question. As \textit{Evaluate} is a judgment type task, more effort is spent on finding support or proof for the particular viewpoint of participant providing answer. Hence, this might have led user to fire fewer number of distinct queries.
To summarize, \textit{Remember} \textless \textit{Understand} \textless \textit{Apply} \textless \textit{Analyze} \textless \textit{Create} and \textit{Remember} \textless \textit{Evaluate} \textless \textit{Create}. Therefore, these results partially support \textit{Hypothesis 2.1}.

\ul{\textit{Hypothesis 2.2:} Query length will increase with the increase in cognitive learning complexity of the task}

\begin{figure}
\centering
\begin{subfigure}{0.5\textwidth}
  \centering
  \includegraphics[width=0.9\linewidth]{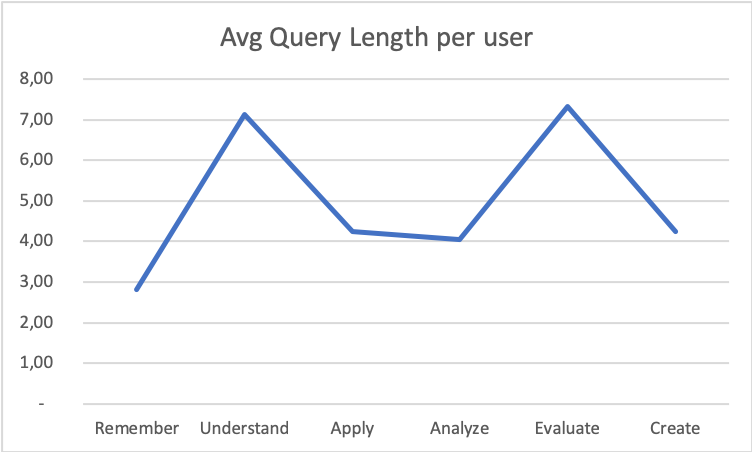}
  \caption{Average query length}
  \label{fig:QueryLenForALLQues}
\end{subfigure}%
\begin{subfigure}{0.5\textwidth}
  \centering
  \includegraphics[width=0.9\linewidth]{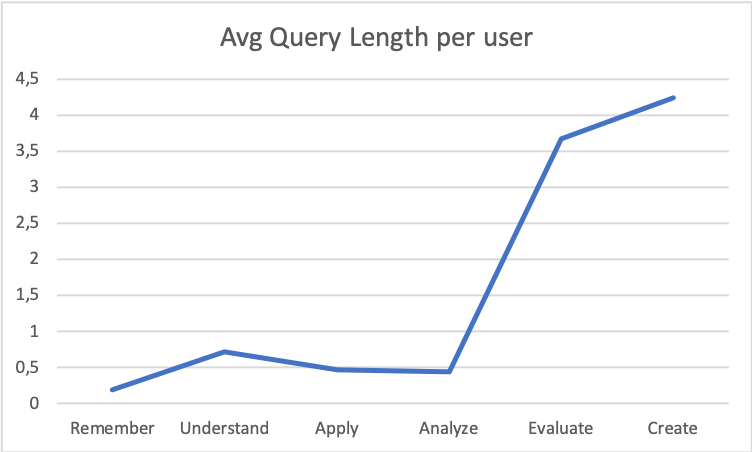}
  \caption{Query length for each user normalized by number of questions}
  \label{fig:queryLenPerQues}
\end{subfigure}
\caption{Average query length for cognitive learning levels}
\label{fig:queryLength}
\end{figure}

A one-way between subjects ANOVA showed significant difference[$F(6, 150) = 47.44, p<0.001$] for the query length across all cognitive levels for user data normalized by the number of questions in each search task. Figure \ref{fig:queryLength} compares the query length trend for both normalized and non-normalized data. While figure \ref{fig:QueryLenForALLQues}, figure using non-normalized user data, does not support \textit{Hypothesis} 2.2, figure \ref{fig:queryLenPerQues} supports \textit{Hypothesis} 2.2 partially. Figure \ref{fig:queryLenPerQues} shows an upward trend in increase in query length from \textit{Remember} to \textit{Create}, however, for the middle tasks - \textit{Understand, Apply,} and \textit{Analyze}, the average query length decreases slightly. So it can be summarized as \textit{Remember} \textless \textit{Understand, Apply, Analyze} \textless \textit{Evaluate} \textless \textit{Create} and \textit{Understand} \textgreater \textit{Apply} \textgreater \textit{Analyze}.

\begin{figure}
\centering
\begin{subfigure}{0.5\textwidth}
  \centering
  \includegraphics[width=0.9\linewidth]{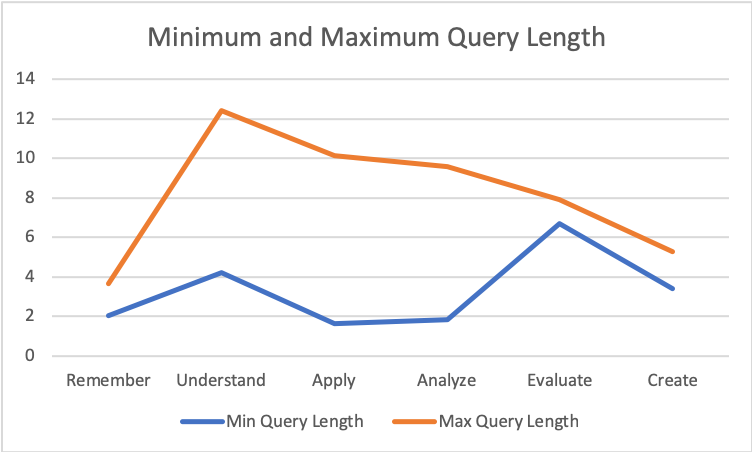}
  \caption{Minimum and maximum query lengths}
  \label{fig:minMaxQueryLengthALLQues}
\end{subfigure}%
\begin{subfigure}{0.5\textwidth}
  \centering
  \includegraphics[width=0.9\linewidth]{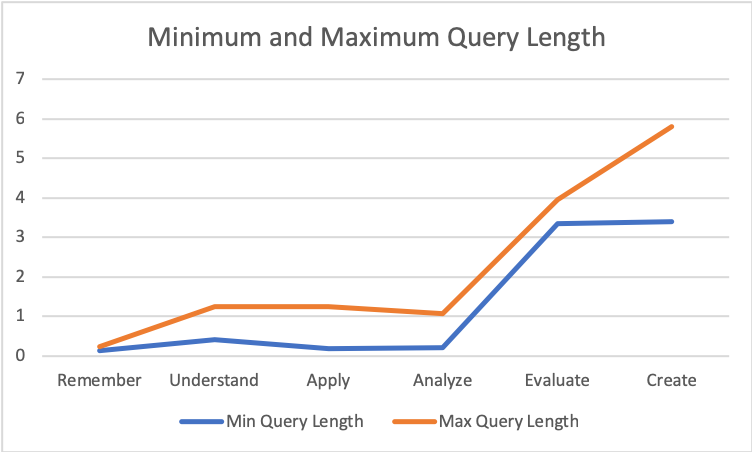}
  \caption{Minimum and maximum query lengths for data normalized by number of questions}
  \label{fig:minMaxQueryLengthPerQues}
\end{subfigure}
\caption{Minimum and maximum query lengths for cognitive learning levels}
\label{fig:minMaxQueryLength}
\end{figure}

\begin{figure}
\centering
\begin{subfigure}{0.5\textwidth}
  \centering
  \includegraphics[width=0.9\linewidth]{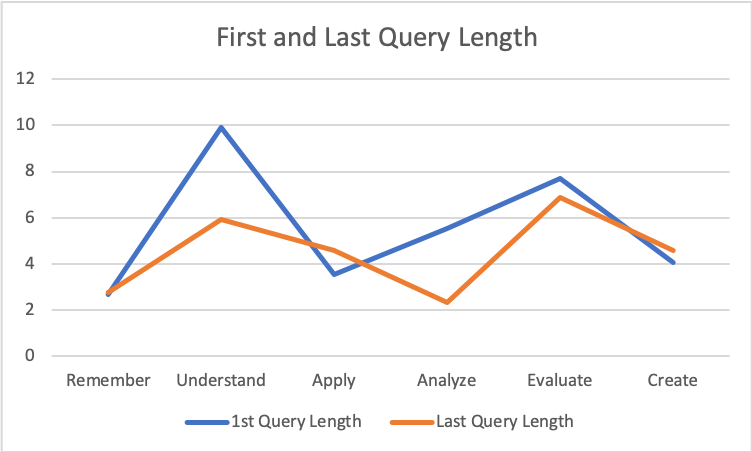}
  \caption{First and last query lengths}
  \label{fig:FLqueryLengthForALLQues}
\end{subfigure}%
\begin{subfigure}{0.5\textwidth}
  \centering
  \includegraphics[width=0.9\linewidth]{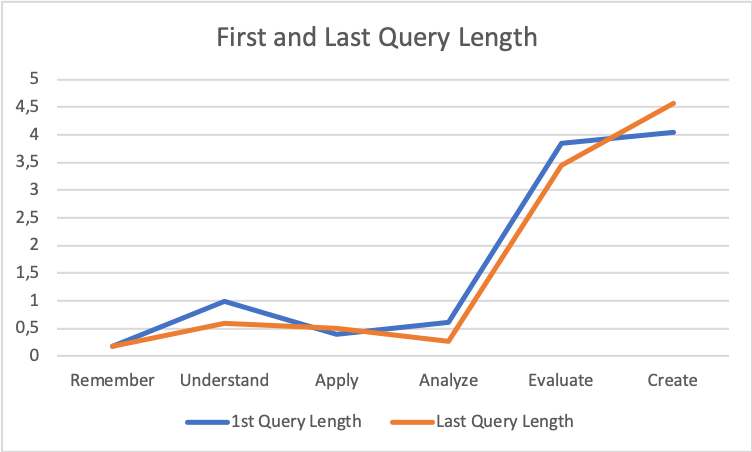}
  \caption{First and last query lengths for data normalized by number of questions}
  \label{fig:FLqueryLengthPerQues}
\end{subfigure}
\caption{First and last query lengths for cognitive learning levels}
\label{fig:FLqueryLength}
\end{figure}

We also, analyzed the first and last query lengths as well as minimum and maximum query lengths. For all of these features, one-way between subjects ANOVA results supported an affect of cognitive learning level on the features for data normalized by number of questions in each search task. Figure \ref{fig:minMaxQueryLength} shows the changes in measurements of minimum and maximum query lengths over cognitive learning levels. Figure \ref{fig:minMaxQueryLengthALLQues} shows no particular trend. However, figure \ref{fig:minMaxQueryLengthPerQues} which refers to minimum and maximum query lengths of user data normalized by number of questions shows an upward graph from lowest complexity to highest complexity, much like the increasing trend of average query length of normalized user data. The same is true for first and last query length as seen in \ref{fig:FLqueryLength}. For maximum query length and last query length the graph for normalized data shows \textit{Remember} \textless \textit{Understand, Apply, Analyze} \textless \textit{Evaluate} \textless \textit{Create} and \textit{Analyze} \textless \textit{Apply} \textless \textit{Understand}. For minimum query length and first query length trend shows \textit{Remember} \textless \textit{Understand, Apply, Analyze} \textless \textit{Evaluate} \textless \textit{Create}, \textit{Apply} \textless \textit{Analyze}, and \textit{Apply} \textless \textit{Analyze} \textless \textit{Understand}. These results too partially support Hypothesis 2.2 as there is an upwards trend in the increase in query search behavior while one goes from a lower complexity task to higher, however, it does not hold true for the search tasks in the intermediate levels of cognitive learning pyramid of revised Bloom's taxonomy.

\ul{\textit{Hypothesis 2.3:} Number of unique terms will increase in number with the increase in cognitive learning complexity of the task}

\begin{figure}
\centering
\begin{subfigure}{0.5\textwidth}
  \centering
  \includegraphics[width=0.9\linewidth]{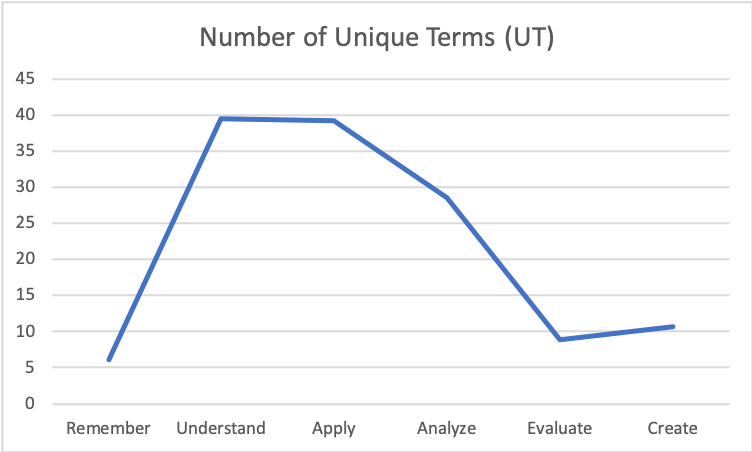}
  \caption{Average unique terms per User}
  \label{fig:UTALLQues}
\end{subfigure}%
\begin{subfigure}{0.5\textwidth}
  \centering
  \includegraphics[width=0.9\linewidth]{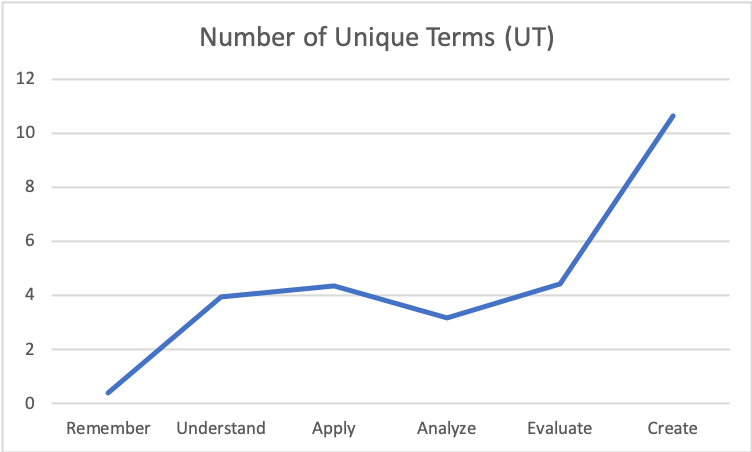}
  \caption{Average unique terms per user from data normalized by number of questions}
  \label{fig:UTPerQues}
\end{subfigure}
\caption{Average number of unique terms across various cognitive learning levels}
\label{fig:UT}
\end{figure}

Analysis of the user interaction data shows that there is a significant effect on number of unique terms in query by the cognitive complexity of the task. While there were many users who saw search queries with only 1 unique term, by the completion of study, the user with maximum number of unique terms in the entire session had carried out search queries including a total of 100 unique terms. Figure \ref{fig:UTALLQues} shows the relation between the average number of unique terms per user and the cognitive learning level of the search task. While this figure does not show any significant pattern in changes in the total number of unique terms across cognitive levels, figure \ref{fig:UTPerQues} which shows the pattern of changes in unique term over a normalized data across cognitive levels based on the number of questions in each search task, shows an upward growth. It indicates that the total number of unique terms per question are lowest for the search task of lowest complexity and highest for the most complex search task. However, the search tasks in intermediate levels of revised Bloom's taxonomy do not exhibit this pattern. It can be seen from the results that \textit{Remember} \textless \textit{Understand, Analyze} \textless \textit{Apply} \textless \textit{Evaluate} \textless \textit{Create} and \textit{Analyze} \textless \textit{Understand}. A one-way between subjects ANOVA for normalized user data supported  that cognitive complexity of the task affects the number of unique terms per question of the task[$F(6, 150) = 24.59, p<0.001$] and that there are significant differences in the results across all the cognitive learning levels.  These results support Hypothesis 2.3 partially.

\begin{figure}
\centering
\begin{subfigure}{0.5\textwidth}
  \centering
  \includegraphics[width=0.9\linewidth]{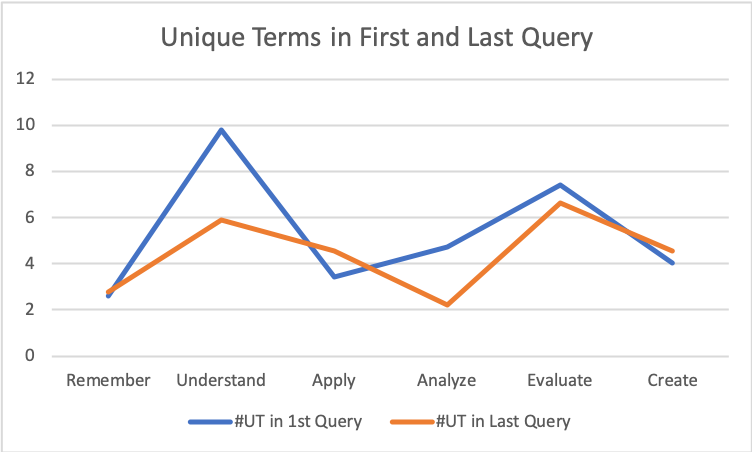}
  \caption{Unique terms in first and last queries}
  \label{fig:UTFLALLQues}
\end{subfigure}%
\begin{subfigure}{0.5\textwidth}
  \centering
  \includegraphics[width=0.9\linewidth]{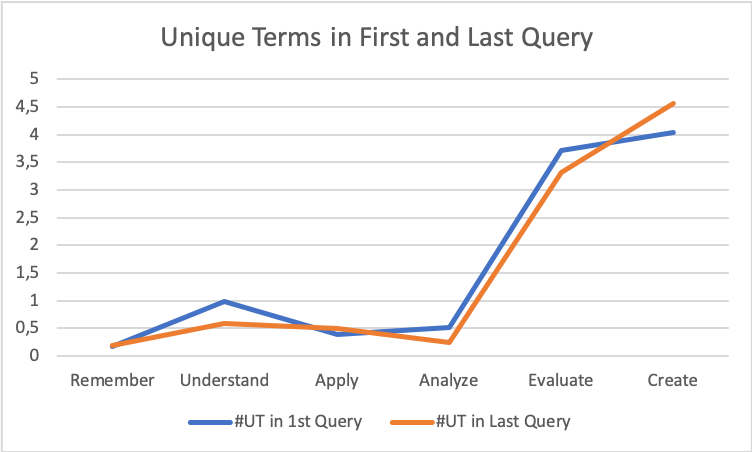}
  \caption{Unique terms in first and last queries from data normalized by number of questions}
  \label{fig:UTFLPerQues}
\end{subfigure}
\caption{Unique number of terms in first and last query}
\label{fig:UTFL}
\end{figure}

We also analyzed the user data normalized by number of questions within task for the number of unique terms in first and last query of users and a one-way between subjects ANOVA showed that the differences in number of unique terms in first query for different cognitive learning levels are statistically significant. The same holds true for the number of unique terms in last query across varying cognitive learning levels. Figure \ref{fig:UTFL} shows the relation between this data and cognitive learning level of search task. As expected, figure \ref{fig:UTALLQues} while showing a significant difference in the data, does not show a significant pattern, however, figure \ref{fig:UTFLPerQues} shows the similar upward trend as seen until now where the behavior observation is more for \textit{Create} which is most complex task and least for \textit{Remember}. The trend for number of unique terms in first query is \textit{Remember} \textless \textit{Understand, Apply, Analyze, Evaluate} \textless \textit{Create} and \textit{Apply} \textless \textit{Analyze} \textless \textit{Evaluate} \textless \textit{Understand}. For the number of unique terms in last query, the trend is \textit{Remember} \textless \textit{Understand, Apply, Analyze} \textless \textit{Evaluate} \textless \textit{Create} and \textit{Analyze} \textless \textit{Apply} \textless \textit{Understand}. These results partially support Hypothesis 2.3
 
\subsection{Websites and Search Pages}

\ul{\textit{Hypothesis 2.4}: Number of websites visited will increase with the increase in cognitive learning complexity  of the task}

\begin{figure}
\centering
\begin{subfigure}{0.5\textwidth}
  \centering
  \includegraphics[width=0.9\linewidth]{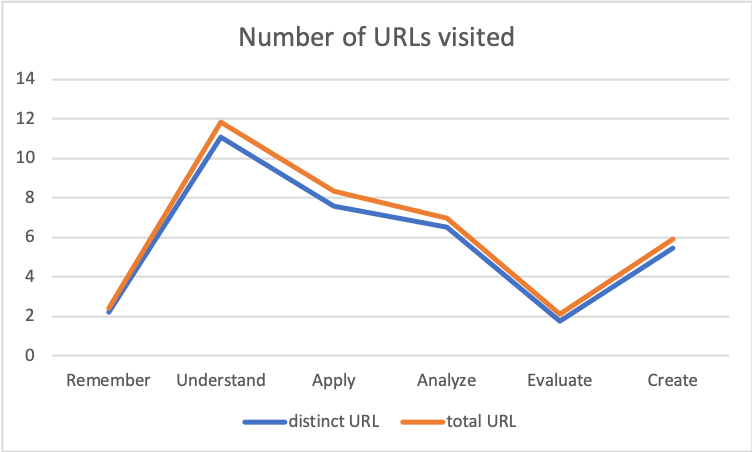}
  \caption{Number of URLs visited}
  \label{fig:URLALLQues}
\end{subfigure}%
\begin{subfigure}{0.5\textwidth}
  \centering
  \includegraphics[width=0.9\linewidth]{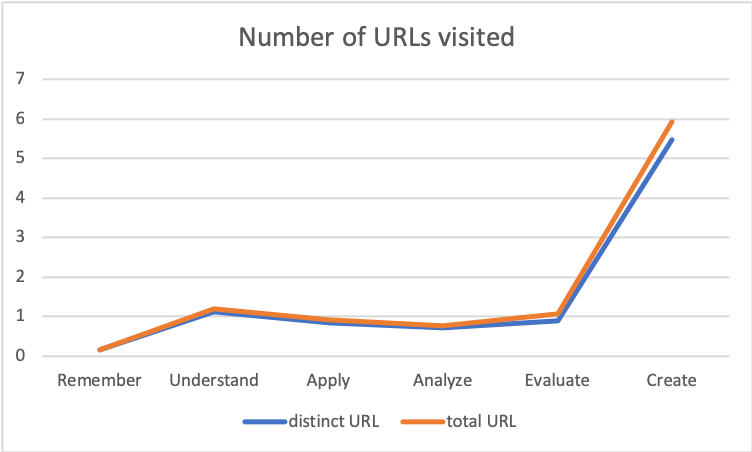}
  \caption{Number of URLs visited for data normalized by number of questions}
  \label{fig:URLPerQues}
\end{subfigure}
\caption{Total and distinct number of URLs visited}
\label{fig:URL}
\end{figure}

User interactions varied from having an average between 2 to 5 web pages visited depending upon the complexity level of the search task to a maximum of 42 pages visited in the entire search session. Figure \ref{fig:URL} shows a plot between total number of web pages visited per user and cognitive learning level. Figure \ref{fig:URLALLQues} shows that while the results are different across cognitive learning levels, the data does not follow the trend set by Hypothesis 2.4. Figure \ref{fig:URLPerQues} shows the distribution of changes in the number of distinct as well as total URLs visited for the data that is normalized by the number of questions in each search task. A one-way between subjects ANOVA for the normalized data supported the presence of a statistically significant difference between the total number of web pages visited and the cognitive learning level of the search task[$F(6, 150)=12.49, p<0.001$] which implies that the total number of web-pages visited by a user is affected by the cognitive learning level of task. The trend for total number of web pages visited normalized by the number of questions within the cognitive learning level is: \textit{Remember} \textless \textit{Understand, Apply, Analyze} \textless \textit{Evaluate} \textless \textit{Create} and, \textit{Analyze} \textless \textit{Apply} \textless \textit{Understand}. 

We also analyzed the number of distinct URLs across varying cognitive levels. A one-way between subjects ANOVA for user data normalized by number of questions within each search task supported that the number of distinct URLs visited by a user is affected by the task complexity[$F(6, 150)=11.33, p<0.001$]. The trends observed are similar to those observed in relation between total number of URLs and cognitive learning level of task.

The above results partially support Hypothesis 2.4.

\ul{\textit{Hypothesis 2.5}: Number of search pages visited will increase with the increase in cognitive learning complexity of the task}

\begin{figure}
\centering
\begin{subfigure}{0.5\textwidth}
  \centering
  \includegraphics[width=0.9\linewidth]{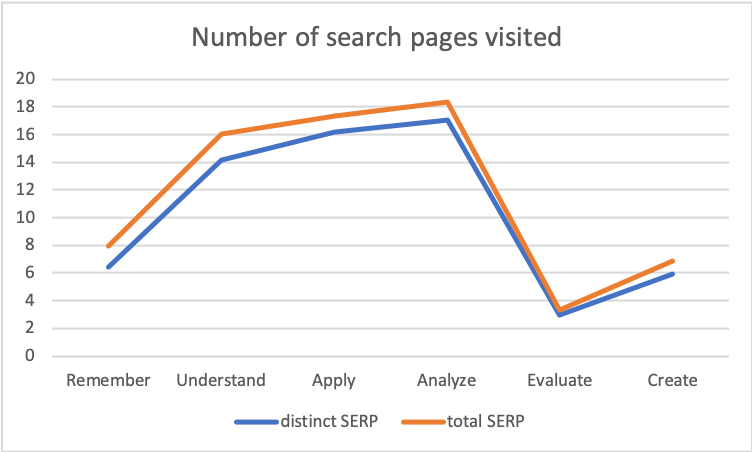}
  \caption{Number of SERP visited}
  \label{fig:SERPALLQues}
\end{subfigure}%
\begin{subfigure}{0.5\textwidth}
  \centering
  \includegraphics[width=0.9\linewidth]{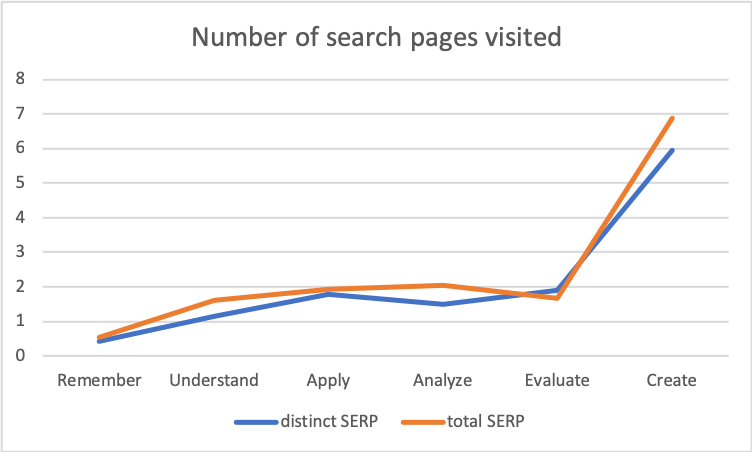}
  \caption{Number of SERP visited for data normalized by number of questions}
  \label{fig:SERPPerQues}
\end{subfigure}
\caption{Total and distinct number of search pages visited across varying cognitive learning levels}
\label{fig:SERP}
\end{figure}

We analyzed the search engine results pages(SERP) consumed by users in search tasks for varying cognitive learning levels. Figure \ref{fig:SERP} illustrates this analysis. Users navigated an average of 3 to 17 distinct search engine result pages during the entire search session. Maximum number of search pages visited by a user was observed to be 39 while attempting \textit{Remember} task. A one-way between subjects ANOVA for user data normalized by the number of task questions showed that the total number of search pages visited is affected by the search complexity of the task[$F(6, 150)=20.87, p<0.001$]. Much like all the other results, the normalized data shows an upward trend where \textit{Remember} \textless \textit{Understand, Evaluate} \textless \textit{Apply} \textless \textit{Analyze} \textless \textit{Create} and \textit{Understand} \textless \textit{Evaluate}. The trend supports the fact that the search behavior for total number of search pages visited is maximum for the highest complexity task and least for lowest complexity task, however, it does not follow the sequence for intermediate search tasks. Hence, it supports Hypothesis 2.5 partially.

Our analysis of distinct number of SERP consumed by user showed similar trend as that of total number of SERP with slight difference in the intermediate levels for the normalized data. A one-way between subjects ANOVA showed that the distinct number of search pages visited too is affected by the search complexity of the task[$F(6, 150)=21.95, p<0.001$] for user data that is normalized by the number of questions in each search task. The trend for the normalized user data is \textit{Remember} \textless \textit{Understand, Analyze} \textless \textit{Apply} \textless \textit{Evaluate} \textless \textit{Create} and \textit{Understand} \textless \textit{Analyze}.

The above results support Hypothesis 2.5 partially.

\subsection{Time Spent Online}

In this section, we will examine the amount of time that the user spent online. This includes the total time required to complete the task as well as the active time spent during the search sessions. Active time is the time in which the user actively carried out interactions in the search session.

\ul{\textit{Hypothesis 2.6:} Time spent online will increase with the increase in cognitive complexity of the task}

\begin{figure}
\centering
\begin{subfigure}{0.5\textwidth}
  \centering
  \includegraphics[width=0.9\linewidth]{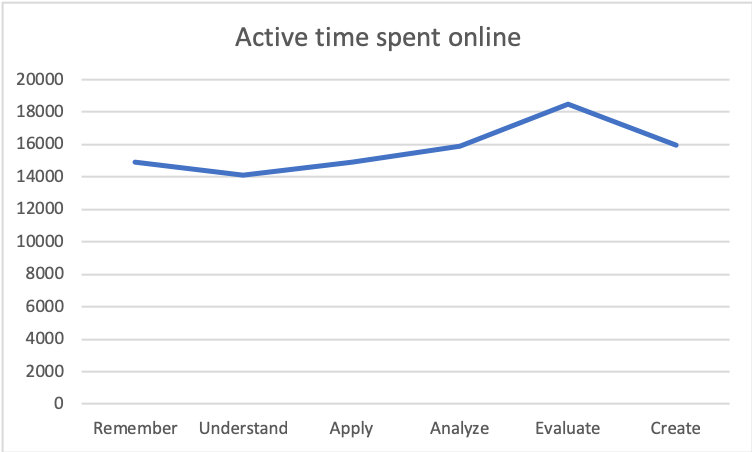}
  \caption{Active time spent by users(in seconds)}
  \label{fig:ActiveTALLQues}
\end{subfigure}%
\begin{subfigure}{0.5\textwidth}
  \centering
  \includegraphics[width=0.9\linewidth]{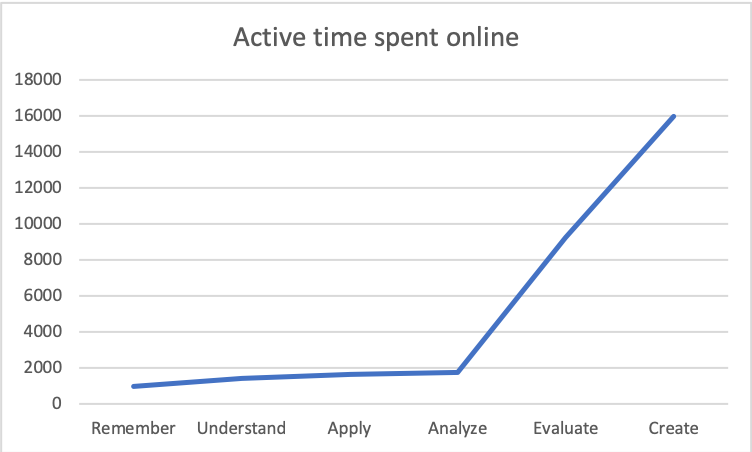}
  \caption{Active time spent by users per question on average(in seconds)}
  \label{fig:ActiveTPerQues}
\end{subfigure}
\caption{Active time spent in seconds across varying cognitive levels}
\label{fig:ActiveTime}
\end{figure}

Time spent while user interacted with the search engine was logged. This time, called user's active time spent on web across all the cognitive learning levels was measured. When the user's active time is normalized by the number of questions in each search task to find the relation between active time spent on average per question and the search complexity of the task; one-way between subjects ANOVA shows that there is significant statistical difference[$F(6, 150)=42.67, p<0.001$] between the active time and the cognitive complexities of the task. This implies that the active time spent online is impacted by the cognitive level of search task. Figure \ref{fig:ActiveTime} shows the relation between active time and cognitive complexity of the task for total active time spent and active time spent per question for varying cognitive levels. Figure \ref{fig:ActiveTALLQues} gives an almost horizontal graph which does not support the hypothesis. However, figure \ref{fig:ActiveTPerQues} gives an upwards trend where \textit{Remember} \textless \textit{Understand} \textless \textit{Apply} \textless \textit{Analyze} \textless \textit{Evaluate} \textless \textit{Create}. Hence, hypothesis 2.6 is supported by these results.

\begin{figure}
\centering
\begin{subfigure}{0.5\textwidth}
  \centering
  \includegraphics[width=0.9\linewidth]{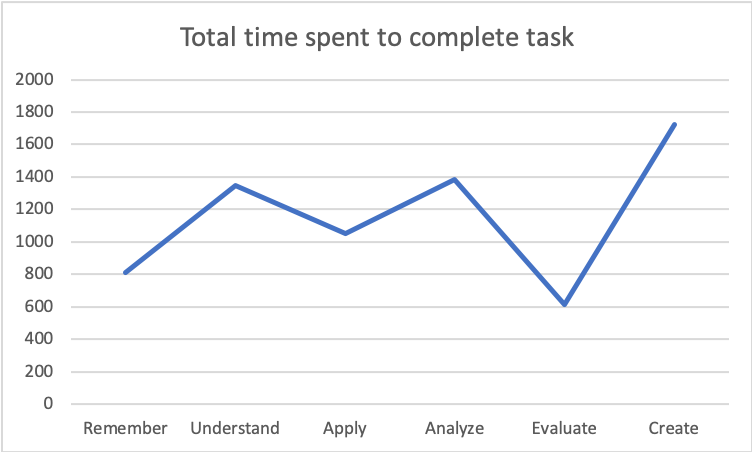}
  \caption{Time spent by users to complete task}
  \label{fig:TotalTALLQues}
\end{subfigure}%
\begin{subfigure}{0.5\textwidth}
  \centering
  \includegraphics[width=0.9\linewidth]{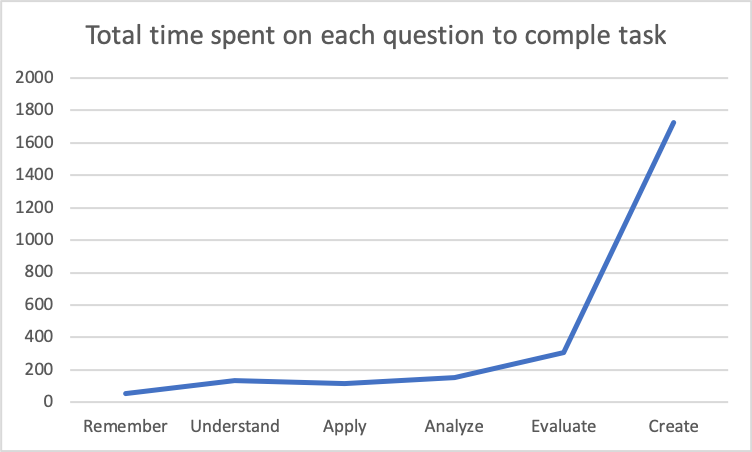}
  \caption{Time spent by users to complete per question on average in task}
  \label{fig:TotalTPerQues}
\end{subfigure}
\caption{Total task duration}
\label{fig:TotalTime}
\end{figure}

We also measured the time taken by user from the moment the task link is opened till the time when the user hits the submit button. We called this duration as the total task duration and compared the total time taken to complete various cognitive levels. Figure \ref{fig:TotalTime} shows the result. However, figure \ref{fig:TotalTALLQues} does not show any significant pattern in between total task duration and cognitive learning level. We also analyzed the total task duration by normalizing the data with the number of questions. The resulting data will give an average amount of time spent in completion of each question in the various search tasks. A one-way between subjects ANOVA supports[$F(6, 150)=10.55, p<0.001$] that the total task duration is affected by the cognitive complexity of the task for user data normalized by the number of questions within each search task. Figure \ref{fig:TotalTPerQues} plots this relationship. It shows an upward trend in time spent much like all the other results seen so far. The trend for total task duration for normalized results is \textit{Remember} \textless \textit{Understand, Apply} \textless \textit{Analyze} \textless \textit{Evaluate} \textless \textit{Create} and \textit{Apply} \textless \textit{Understand}. Hence, these results partially support Hypothesis 2.6

Results stated to support Hypothesis 2.1, 2.2, 2.3, 2.4, and 2.6 show that Hypothesis 2 is partially supported. This means that search behavior in the form of user interactions is minimum for lowest cognitive complexity level and highest for maximum complexity level. However, the increase in search behavior does not follow the taxonomic structure for intermediate levels.



Following table \ref{tab:Trend} summarizes the results for each of the search behavior property. The table outlines the trend observed for each search task. Table \ref{tab:ANOVA} summarizes the one-way between subjects ANOVA results for each of the search behavior property. We publish the entire user data for public at \href{https://github.com/rishitakalyani/userInteractions}{https://github.com/rishitakalyani/userInteractions}.

\begin{longtable}{L{2.5cm}C{5cm}C{6cm}}
    \hline
         \textbf{Measuring feature} & \textbf{Data normalized by Question No} & \textbf{Data averaged by User No} \\\midrule\midrule
         Knowledge Gain\% & \textit{Understand \textgreater Remember \textgreater Understand \textgreater Analyze}  & \textit{Remember \textgreater Understand, Apply \textgreater Analyze and,\newline Apply \textgreater Understand} \\\midrule
         No. of Distinct Queries & \textit{Remember, Evaluate \textless Understand \textless Apply \textless Analyze \textless Create and,\newline Remember \textless Evaluate} & \textit{Remember \textless Understand \textless Apply, Analyze and, \newline Evaluate \textless Remember \textless Create \textless Analyze \textless Apply}\\\midrule
         Average Query Length(QL) & \textit{Remember \textless Understand, Apply, Analyze \textless Evaluate \textless Create and, \newline Analyze \textless Apply \textless Understand} & \textit{Remember \textless Understand \textless Evaluate and, \newline Remember \textless Apply, Analyze \textless Create}\\\midrule
         Min. QL &  \textit{Remember \textless Understand, Apply, Analyze \textless Evaluate \textless Create and, \newline Apply \textless Analyze \textless Understand} & \textit{Remember \textless Understand \textless Evaluate and, \newline Apply \textless Analyze \textless Create}\\\midrule 
         Max. QL &  \textit{Remember \textless Understand, Apply, Analyze \textless Evaluate \textless Create and, \newline Analyze \textless Apply, Understand} & \textit{Remember, Create \textless Evaluate \textless Analyze \textless Apply \textless Understand and, \newline Remember \textless Create}\\\midrule 
         1\textsuperscript{st} QL &  \textit{Remember \textless Understand, Apply, Analyze \textless Evaluate \textless Create and, \newline Apply \textless Analyze \textless Understand} & \textit{Remember \textless Apply \textless Create  \textless Analyze \textless Evaluate  \textless Understand} \newline $\implies$ no significant pattern \\\midrule 
         Last QL &  \textit{Remember \textless Understand, Apply, Analyze \textless Evaluate \textless Create and, \newline Analyze \textless Apply \textless Understand} & \textit{ Analyze \textless Remember \textless Apply, Create  \textless Understand \textless Evaluate } \newline $\implies$ no significant pattern \\\midrule 
         Unique Terms(UT) & \textit{Remember \textless Understand \textless Apply, Analyze \textless Evaluate \textless Create and, \newline Analyze \textless Apply} & \textit{Remember \textless Understand and, \newline Create, Evaluate \textless Analyze \textless Apply \textless Understand}\\\midrule
         UT in 1\textsuperscript{st} Query & \textit{Remember \textless Understand, Apply, Analyze \textless Evaluate \textless Create and, \newline Apply \textless Analyze \textless Understand} & \textit{Remember \textless Apply \textless Create \textless Analyze \textless Evaluate \textless Understand \newline $\implies$ no significant pattern}\\\midrule
         UT in Last Query & \textit{Remember \textless Understand, Apply, Analyze \textless Evaluate \textless Create and, \newline Analyze \textless Apply \textless Understand} & \textit{Analyze \textless Remember \textless Create, Apply \textless Understand \textless Evaluate \newline $\implies$ no significant pattern}\\\midrule
         Total URLs & \textit{Remember \textless Understand, Apply, Analyze, Evaluate \textless Create and, \newline Analyze \textless Apply \textless Evaluate \textless Understand} & \textit{Evaluate \textless Remember \textless Create \textless Analyze \textless Apply \textless Understand  \newline $\implies$ no significant pattern}\\\midrule
         Distinct URLs & \textit{Remember \textless Understand, Apply, Analyze, Evaluate \textless Create and, \newline Analyze \textless Apply \textless Evaluate \textless Understand} & \textit{Evaluate \textless Remember \textless Create \textless Analyze \textless Apply \textless Understand  \newline $\implies$ no significant pattern}\\\midrule
         Total SERP & \textit{Remember \textless Understand \textless Apply, Evaluate \textless Analyze \textless Create and, \newline Apply \textless Evaluate} & \textit{Remember \textless Understand \textless Apply \textless Analyze and, \newline Evaluate \textless Create \textless Analyze } \\\midrule
         Distinct SERP & \textit{Remember \textless Understand \textless Apply, Analyze \textless Create and, \newline Analyze \textless Apply} & \textit{Remember \textless Understand \textless Apply \textless Analyze and, \newline Evaluate \textless Create \textless Analyze } \\\midrule
         Active Time &\textit{Remember \textless Understand \textless Apply \textless Analyze \textless Create} & \textit{Remember $\sim$ Understand $\sim$ Apply \textless Analyze $\sim$ Create \textless Evaluate} \\\midrule
         Task duration &\textit{Remember \textless Understand, Apply, Analyze \textless Evaluate \textless Create and, \newline Apply \textless Analyze \textless Understand} & \textit{Remember, Understand, Apply \textless Analyze \textless Create and, \newline Apply \textless Remember \textless Understand} \\\midrule
         \bottomrule
    \caption{Trend summary for each feature of user behavior and the data collected from user interaction}
    \label{tab:Trend}
\end{longtable}

\begin{longtable}{L{4cm}C{6cm}}
    \hline
         \textbf{Measuring feature} & \textbf{Data normalized by Question No}  \\\midrule\midrule
         Knowledge Gain\% & $F(6,150) =21.88$\newline $p<0.001$ \newline true \\\midrule
         No. of Distinct Queries & $F(6,150) =27.68$\newline $p<0.001$ \newline true \\\midrule
         Average Query Length(QL) & $F(6,150) =47.44$\newline $p<0.001$ \newline true\\\midrule
         Min. QL & $F(6,150) =12.68$\newline $p<0.001$ \newline true \\\midrule 
         Max. QL &   $F(6,150) =15.10$\newline $p<0.001$ \newline true \\\midrule 
         1\textsuperscript{st} QL &  $F(6,150) =24.12$\newline $p<0.001$ \newline true \\\midrule 
         Last QL & $F(6,150) =16.18$\newline $p<0.001$ \newline true \\\midrule 
         Unique Terms(UT) &$F(6,150) =24.59$\newline $p<0.001$ \newline true \\\midrule
         UT in 1\textsuperscript{st} Query & $F(6,150) =25.77$\newline $p<0.001$ \newline true\\\midrule
         UT in Last Query & $F(6,150) =16.67$\newline $p<0.001$ \newline true \\\midrule
         Total URLs &  $F(6,150) =12.49$\newline $p<0.001$ \newline true\\\midrule
         Distinct URLs &  $F(6,150) =11.33$\newline $p<0.001$ \newline true \\\midrule
         Total SERP &  $F(6,150) =20.87$\newline $p<0.001$ \newline true \\\midrule
         Distinct SERP &  $F(6,150) =21.95$\newline $p<0.001$ \newline true\\\midrule
         Active Time &  $F(6,150) =42.67$\newline $p<0.001$ \newline true  \\\midrule
         Task duration &  $F(6,150) =10.55$\newline $p<0.001$ \newline true \\\midrule
         \bottomrule
    \caption{One-way between subjects ANOVA for each feature of user behavior and the data collected from user interaction}
    \label{tab:ANOVA}
\end{longtable}

\section{Discussion of Results and Limitations}
We designed a study based on the revised Bloom's taxonomic structure and explored how learning related search tasks where each search task corresponds to a unique cognitive learning level affects the knowledge that is gained as well as the search behavior.

We aimed at finding if our study setup of disjointing each cognitive learning search task from other would help us identify the the impact of search task on user's activity and increase in knowledge.

First, in response to our Research Question 1, we were able to prove that knowledge is changed by the search tasks of first four cognitive levels empirically. We also established that knowledge gain is impacted by the cognitive complexity for first four cognitive learning levels by one-way between subjects ANOVA. However, we weren't able to find a significant pattern between knowledge gain and the hierarchy of complexity level, i.e., we weren't able to show that knowledge gain either increases or decreases as complexity of task increases. One of the reasons why we were unsuccessful in finding a trend between knowledge gain and cognitive learning complexity was because we limited our calculation of knowledge gain only up-to first four levels. As it is seen from the search behavior pattern, the results for intermediate levels does not always follow a successive trend. If this is reflected for knowledge gain as well, it makes it difficult for us to establish a trend in changes in knowledge gain across the cognitive complexity levels without empirical details for the final two cognitive levels. Finally, we also believe that the behavior of knowledge change can be presumed to have taken place for \textit{Evaluate} and \textit{Create} tasks as well since these two tasks have users interacting with web in similar patterns as the lower four cognitive level, and upon manual checking of answers, the users who carried web research before submitting answers, submitted valid entries.

In answer to the second research question, we found that various user interactions across the cognitive learning level had statistically significant impact by the cognitive level of search task on user interactions. Further, for the normalized data user interactions followed an upward trend for from \textit{Remember} to \textit{Create} for all of the observed user interactions, i.e., the search behavior in terms of user interactions increased as one carried more complex task. While this was an expected result for second research question as it was assumed that when user would carry a more complex task he will require more effort in information-finding, ours is the only study that has been able to detect this pattern for all user interactions until now. We attribute this finding to our unique way of designing the the set-up for the crowd-sourcing experiments. The results also showed that while trend was upwards, the intermediate tasks did not strictly follow the trend. This meant that on few occasions, user interactions were observed more for an immediately lower complexity level than for the current one. However, the difference observed was never an abnormally large one in almost all of the cases. Hence, we attribute these anomalies to the fact that \citeauthor{krathwohl2002revision}'s revised taxonomy is not rigidly bound and the transition from one level to another is a continuous process. Since there are no rigid distinctions between the levels, it is easy for a user interaction feature to fall out of the expected boundary for a particular level.

\subsection{Limitations}

While the results are useful, there are certain limitations to it. Since this study uses crowd-sourced user data, it brings with it the limitations of crowdsourcing platforms like being unable to ensure that the worker is invested in the experimental task\citep{gadiraju2017crowdsourcing}. The results gathered are from crowd-sourced workers who do not always participate in a task with a learning motivation but with a motivation to get paid\citep{han2019all}. Further, the tasks were constrained to a maximum of 30 minute duration and hence, it was not possible for us to determine knowledge gain over a longer stretch of time.

One of the major limitations of the study was being unable to determine the difficulty of each task that was experienced through participating user's assessment of task difficulty. The reason for this is the approach we took in designing our experimental set-up on crowd-sourcing platform. As a user would attempt only one of the tasks corresponding to one cognitive learning complexity, we couldn't ask him to answer how much more or less difficult the attempted task is in comparison to other tasks of the remaining cognitive learning complexities. Hence, it was not possible for us to provide the user data for observed difficulty of the tasks.
 
\chapter{Conclusion and Future Work} 

\label{Chapter6} 

\lhead{Chapter 6. \emph{Conclusion and Future Work}} 


In this thesis, we developed six search tasks corresponding to each of the \citeauthor{krathwohl2002revision}'s cognitive complexity levels. These tasks were designed specifically to measure user interactions and knowledge gain across varying cognitive learning levels. We asked participants of a crowd-sourcing platform to answer the search tasks in a consecutive manner. The unique design of the experimental set-up on the crowd-sourcing platform ensured that no user participated in search tasks of two or more cognitive learning complexity levels. This allowed us to accumulate user's search behavior which was purely for the chosen cognitive complexity. We analyzed the user's search interactions to establish a relation between search behavior and the taxonomic structure of revised Bloom's taxonomy.

Not only did we collect user's search interaction but we also calculated user's knowledge gain by devising a set of rules to compute user's knowledge gain for the first four cognitive domains of revised taxonomic structure. We also demonstrated a relation between the search task complexity and knowledge gain with empirical data that showed an impact in knowledge gain by the cognitive learning level.

The results of user's interaction was able to support the revised Bloom's taxonomic structure with empirical proof for the first time. While not true for each domain level, user's interactions however, showed a general upward trend while traveling from a lowest cognitive complexity level to highest. This allowed us to conclude that if user advances to highest complexity level from lowest, there will definitely be an increase in the observed search behavior.

This study is one of the unique studies that was able to find a statistically significant difference in all of the discussed user interactions across all of the cognitive learning levels of revised Bloom's taxonomy. The impact of this finding reveals the importance of the search tasks designed. The search tasks can be used to study the user interactions by search interfaces facilitating learning needs. It can also be used to used to enhance specific cognitive needs. Further, since a one-way between subjects ANOVA shows that all the results are significantly different across the cognitive levels, a machine learning model can be designed to train the model with the user interaction data collected which will allow in detecting user's cognitive learning level or even predict the possible knowledge gain. We aim to present this supervised machine learning model in near future.

For our future studies we would also like to conduct the same experiments for different topic domain to eliminate any topic bias that could have appeared in the results. We would also to like conduct a study where the 30 minute time limit is relaxed to study the knowledge gain over a longer period to determine the consequences of time on knowledge gain.



\addtocontents{toc}{\vspace{2em}} 

\appendix 



\chapter{Search Task Questions} 

\label{AppendixA} 

\lhead{Appendix A. \emph{Search task Questions}} 

\section{Question Bank - ``Vitamins and Nutrients''}\label{sec:questionBank}

\subsection{Remember}\label{sec:RememberQuestion}
\begin{enumerate}
    \item Rickets is caused by deficiency of vitamin \underline{\hspace{1.5cm}D\hspace{1.5cm}}.
    \item \underline{\hspace{1.5cm}avitaminosis/hypovitaminosis\hspace{1.5cm}} is a disease that is also known as vitamin deficiency.
    \item Night-blindness is caused by deficiency of vitamin \underline{\hspace{1.5cm}A\hspace{1.5cm}}.
    \item \underline{\hspace{1.5cm}anti-vitamins\hspace{1.5cm}} are chemical compounds that inhibit the absorption or actions of vitamins.
    \item Scurvy is caused by deficiency by of vitamin \underline{\hspace{1.5cm}C\hspace{1.5cm}}.
    \item The two types of nutrients are \underline{\hspace{1cm}macronutrients\hspace{1cm}} and \underline{\hspace{0.5cm}micronutrients\hspace{0.5cm}}.
    \item \underline{\hspace{1.5cm}hypervitaminosis\hspace{1.5cm}} is a condition caused by abnormally high storage of vitamins.
    \item Inadequate protein intake can cause \underline{\hspace{1.5cm}kwashiorkor\hspace{1.5cm}} which is also known as edematous malnutrition.
    \item Vitamin \underline{\hspace{1.5cm}B\hspace{1.5cm}} and Vitamin \underline{\hspace{1.5cm}C\hspace{1.5cm}} are water-soluble vitamins.
    \item Vitamins \underline{\hspace{1cm}A\hspace{1cm}}, \underline{\hspace{1cm}D\hspace{1cm}}, \underline{\hspace{1cm}E\hspace{1cm}}, and \underline{\hspace{1cm}K\hspace{1cm}} are fat-soluble vitamins.
    \item Hyponatremia and hypernatremia are terms of defieciency and excess related to \underline{\hspace{1.5cm}sodium\hspace{1.5cm}}.
    \item Recommended intake of vitamin C for an adult according to United States dietary allowance is \underline{\hspace{1.5cm}75-90\hspace{1.5cm}} (mg).
    \item Beriberi is caused by deficiency of vitamin
    \underline{\hspace{1.5cm}Thiamine/B1\hspace{1.5cm}}.
    \item Anaemia is caused by the deficiency of \underline{\hspace{1.5cm}iron\hspace{1.5cm}}.
    \item Overnutrition is a form of malnutrition. True or False? \underline{\hspace{1.5cm}true\hspace{1.5cm}}.
\end{enumerate}

\subsection{Understand}\label{sec:UnderstandQuestion}
\begin{enumerate}
    \item Select the correct sentences from below
    \begin{enumerate}[label=(\alph*)]
        \item \ding{52} An adult's diet may be deficient in vitamins, especially vitamins A and D for many months before they develop deficiency
        \item \ding{54} Excess vitamin B in human body will be stored in liver
        \item \ding{52} Hypervitaminoses are primarily caused by fat-soluble vitamins than water-soluble vitamins
        \item \ding{54} Two men, one with dark skin-tone and another with light skin-tone, exposed to abundant sunlight are equally prone to having vitamin d deficiency
        \item \ding{54} Anti-vitamins are compounds that remove vitamins from body
    \end{enumerate}
    \item What are the common symptoms associated with vitamin D toxicity
    \begin{enumerate}[label=(\alph*)]
        \item \ding{52} Vomiting
        \item \ding{52} Fatigue
        \item \ding{54} Increased appetite
        \item \ding{54} Diarrhea
        \item \ding{52} Muscle weakness
    \end{enumerate}
    \item What are the common symptoms associated with avitaminosis C?
    \begin{enumerate}[label=(\alph*)]
        \item \ding{52} Bleeding gums
        \item \ding{54} Skin rashes
        \item \ding{52} Weakness
        \item \ding{54} Blurry vision
        \item \ding{52} Depression
    \end{enumerate}
     \item Select food sources of Thiamine vitamin 
    \begin{enumerate}[label=(\alph*)]
        \item \ding{54} Soy milk
        \item \ding{52} eggs
        \item \ding{54} Nuts
        \item \ding{52} Pork
        \item \ding{54} yellow fruits
    \end{enumerate}
    \item Which of the following are water-soluble vitamins?
    \begin{enumerate}[label=(\alph*)]
        \item \ding{54} Vitamin D
        \item \ding{52} Vitamin B$_2$
        \item \ding{52} Vitamin C
        \item \ding{54} Vitamin A
        \item \ding{52} Vitamin B$_6$
    \end{enumerate}
    \item Vitamin A deficiency can cause which of the following diseases?
    \begin{enumerate}[label=(\alph*)]
        \item \ding{54} Osteomalacia
        \item \ding{52} Night blindness
        \item \ding{54} Anemia 
        \item \ding{52} Keratomalacia
        \item \ding{54} Pellagra
    \end{enumerate}
    \item Which of the following are one of the B-Complex vitamins?
    \begin{enumerate}[label=(\alph*)]
        \item \ding{52} Biotin
        \item \ding{52} Folates
        \item \ding{54} Tocopherols
        \item \ding{52} Pantothenic acid
        \item \ding{54} Beta carotene
    \end{enumerate}
    \item Eggs are rich in which of the following vitamins?
    \begin{enumerate}[label=(\alph*)]
        \item \ding{54} Vitamin A
        \item \ding{52} Vitamin B
        \item \ding{54} Vitamin C
        \item \ding{52} Vitamin D
        \item \ding{54} Vitamin E
    \end{enumerate}
    \item Select the incorrect statements from below
    \begin{enumerate}[label=(\alph*)]
        \item \ding{54} Provitamin is a substance that may be converted within the body to a vitamin
        \item \ding{52} Vitamin D deficiency can occur only in kids and infants
        \item \ding{52} A human having a diet rich in vitamins such that there will be no vitamin toxicity as well, will never have a vitamin deficiency
        \item \ding{54} Humans need vitamins through diet because their bodies cannot synthesize them naturally
        \item \ding{54} Each vitamin has multiple functions in a body
    \end{enumerate}
    \item Select food sources of Vitamin A
    \begin{enumerate}[label=(\alph*)]
        \item \ding{54} Banana
        \item \ding{52} Liver
        \item \ding{54} Eggs
        \item \ding{52} Orange 
        \item \ding{54} Brown rice
    \end{enumerate}
\end{enumerate}

\subsection{Apply}\label{sec:ApplyQuestion}

\begin{enumerate}
    \item Order the following events in correct sequence
        \begin{enumerate}[label=(\alph*)]
        \item Mary was diagnosed with Scurvy
        \item Mary felt that lemons gave her headaches
        \item Mary started a new diet avoiding lemon
        \item Mary took supplements
        \item A blood test was carried out for Mary
        \item Mary started having bleeding gums, hair loss and lethargic
        \item[]Answer: (b) \ding{212} (c) \ding{212} (f) \ding{212} (e) \ding{212} (a) \ding{212} (d)
    \end{enumerate}
    \item Which of the following are nutrients but not vitamins?
    \begin{enumerate}[label=(\alph*)]
        \item \ding{54} Biotin
        \item \ding{52} Iron
        \item \ding{52} Calcium
        \item \ding{54} Tocopherol
        \item \ding{52} Copper
    \end{enumerate}
    \item Mary frequently gets muscle pain especially in her legs during night. Mary's symptoms are most likely associated with which vitamin deficiency?
    \begin{enumerate}[label=(\alph*)]
        \item \ding{54} Vitamin A
        \item \ding{52} Vitamin B
        \item \ding{54} Vitamin C
        \item \ding{52} Vitamin D
        \item \ding{52} Vitamin E
    \end{enumerate}
    \item Order the following events in correct sequence
        \begin{enumerate}[label=(\alph*)]
            \item Lucy suffers from irritation, hair loss, liver damage, vomiting
            \item Lucy decided to take supplements for nutrients to overcome lack of food 
            \item Lucy started a new diet
            \item Lucy wanted to lose some weight
            \item Lucy took too many supplements
            \item[] Answer: (d) \ding{212} (c) \ding{212} (b) \ding{212} (e) \ding{212} (a)
        \end{enumerate}
    \item Interprete the following symptoms and diagnose the deficiency
    \begin{itemize}
        \item Swelling of gums and gum diseases
        \item Edema after many months
        \item Lethargy
        \item Malaise
        \item Jaundice in late stage
    \end{itemize}
    \begin{enumerate}[label=(\alph*)]
        \item \ding{54} Vitamin A deficiency
        \item \ding{54} Iron deficiency
        \item \ding{52} Vitamin C deficiency
        \item \ding{54} Cobalt deficiency
        \item \ding{54} Vitamin E deficiency
    \end{enumerate}
    \item Mary noticed that her skin is pale. She often has cravings to eat dirt, chalk, coal, etc. She gets weak very often and often needs to wear warm socks and gloves as her hands and feet become cold very easily. What could be the most common cause of Mary's symptoms?
   \begin{enumerate}[label=(\alph*)]
        \item \ding{54} Vitamin E deficiency
        \item \ding{54} Potassium deficiency
        \item \ding{54} Vitamin K deficiency
        \item \ding{54} Thiamin deficiency
        \item \ding{52} Iron deficiency
    \end{enumerate}
    \item Lucy has Rickets, what should she do?
    \begin{enumerate}[label=(\alph*)]
        \item \ding{52} Get a bone density scan
        \item \ding{52} Change her diet
        \item \ding{54} Blood transfusion
        \item \ding{52} Get exposed to more sunlight
        \item \ding{54} Treat with nicotinamide
    \end{enumerate}
    \item To continue a healthy life, it is important to have a full-rounded diet. Including which of the following will make your diet a full-rounded one
    \begin{enumerate}[label=(\alph*)]
        \item \ding{52} Vitamins
        \item \ding{52} Proteins
        \item \ding{52} Minerals
        \item \ding{52} Acids
        \item \ding{54} Carbon dioxide
    \end{enumerate}
    \item Which of the following are necessary for healthy bone growth?
    \begin{enumerate}[label=(\alph*)]
        \item \ding{54} Vitamin B
        \item \ding{54} Vitamin C
        \item \ding{52} Vitamin D
        \item \ding{52} Calcium
        \item \ding{54} Iron
    \end{enumerate}
\end{enumerate}

\subsection{Analyze}\label{sec:AnalyzeQuestion}
\begin{enumerate}
    \item Which of the following statements are true for minerals and which of them are true for vitamins?
    \begin{enumerate}[label=(\alph*)]
        \item They are categorized as micro-nutrients
        \item A few of them are micro-nutrients and a few of them are macro-nutrients
        \item They can not be synthesized by the body
        \item They are chemical compounds
        \item They are organic compounds
    \end{enumerate}
    Statements true for minerals: \underline{\hspace{1.5cm}b, c, and d\hspace{1.5cm}}\newline
    Statements true for vitamins: \underline{\hspace{1.5cm}a, c, and e\hspace{1.5cm}}
    \item Which of the following statements are true for both vitamins and minerals?
    \begin{enumerate}[label=(\alph*)]
        \item \ding{52} If the body is not able to absorb them properly then certain kinds of deficiency may occur
        \item \ding{52} Too much intake of them may lead to toxicity in body
        \item \ding{52} Consuming them through diet in proper amounts may still not ensure that one will not suffer from deficiency
        \item \ding{54} There are chemical compounds which inhibit the absorption or actions of them
        \item \ding{54} They are either water-soluble or fat-soluble
    \end{enumerate}
    \item Which of the following statements are true for vitamins, minerals as well as proteins?
    \begin{enumerate}[label=(\alph*)]
        \item \ding{54} They are fat-soluble
        \item \ding{54} They are building blocks for body
        \item \ding{52} They are ingested through one's diet
        \item \ding{54} There are clear guidelines for the amount of intake of them
        \item \ding{52} They are essential nutrients
    \end{enumerate}
    \item Which of the following statements are true for deficiencies and which of them are true for toxicity?
    \begin{enumerate}[label=(\alph*)]
        \item This occurs when one consumes too much of nutrients
        \item This occurs when one consumes synthetic supplements without proper guidance
        \item This occurs when one's body is not able to absorb nutrients on its own
        \item This is common for fat-soluble vitamins
        \item This can lead to many diseases
    \end{enumerate}
    Statements true for deficiency: \underline{\hspace{1.5cm}c and e\hspace{1.5cm}}\newline
    Statements true for toxicity: \underline{\hspace{1.5cm}a, b, and e\hspace{1.5cm}}
    \item Categorize the following into minerals, vitamins and other nutrients?
    \begin{enumerate}[label=(\alph*)]
        \item Riboflavin
        \item Iron
        \item Phosphorous
        \item Fats
        \item Tocopherols
        \item Carbohydrates
    \end{enumerate}
    Minerals: \underline{\hspace{1.5cm}b and c\hspace{1.5cm}}\newline
    Vitamins: \underline{\hspace{1.5cm}a and e\hspace{1.5cm}}\newline
    Others: \underline{\hspace{1.5cm}d and f\hspace{1.5cm}}
    \item Which of the following statements are true for macro-nutrients and which are true for micro-nutrients?
    \begin{enumerate}[label=(\alph*)]
        \item They provide energy
        \item They are consumed in large quantities
        \item They support metabolism
        \item Fats and carbohydrates are examples of this type of nutrients
        \item They are consumed in small quantities everyday
        \item They are important for a healthy functioning of body and its daily activities
    \end{enumerate}
    Statements true for macro-nutrients: \underline{\hspace{1.5cm}a, b, d, and f\hspace{1.5cm}}\newline
    Statements true for micro-nutrients: \underline{\hspace{1.5cm}c, e, and f\hspace{1.5cm}}
    \item Categorize following into vitamin deficiency or mineral deficiency
    \begin{enumerate}[label=(\alph*)]
        \item Beriberi
        \item Anaemia
        \item Night blindness
        \item Pellagra
        \item Hypocalcaemia
    \end{enumerate}
    Vitamin deficiencies: \underline{\hspace{1.5cm}a, c, and d\hspace{1.5cm}}\newline
    Mineral deficiencies: \underline{\hspace{1.5cm}b and e\hspace{1.5cm}}\newline
    \item Which statements are true for malnutrition and which are true for overnutrition?
    \begin{enumerate}[label=(\alph*)]
        \item It is a condition that results from eating a diet in which one or more nutrients are either not enough or too much such that the it causes health problems
        \item Obesity is one of the causes of this condition
        \item This condition is caused by overeating
        \item If this condition occurs during pregnancy, it may result in permanent problems with physical and mental development
        \item In certain poor countries, due to poverty and not having enough food to eat may lead to extreme hunger
    \end{enumerate}
    Malnutrition: \underline{\hspace{1.5cm}a, b, c, d, and e\hspace{1.5cm}}\newline
    Overnutrition: \underline{\hspace{1.5cm}b and c\hspace{1.5cm}}\newline
    \item Categorize the following into either mineral toxicity or hypovitaminosis
    \begin{enumerate}[label=(\alph*)]
        \item Hypercalcemia 
        \item Hypervitaminosis D
        \item Hypervitaminosis E
        \item Hypermagnesemia
        \item Hemochromatosis
        \item Hypernatremia
    \end{enumerate}
    Mineral toxicity: \underline{\hspace{1.5cm}a, d, e, and f\hspace{1.5cm}}\newline
    Hypovitaminosis: \underline{\hspace{1.5cm}b and c\hspace{1.5cm}}\newline
\end{enumerate}

\subsection{Evaluate}\label{sec:EvaluateQuestion}
Your nephew started a new diet to lose weight and he has stopped eating food in required amount. To overcome the lack in nutrients, he has decided to take supplements. He believes that it is better to take vitamins and nutrients in the form of supplements rather than from fresh food as he can lose weight easily this way.

Do you agree with what your nephew is doing?\newline (Yes/No)\underline{\hspace{3cm}}

Before writing down your justification for above answer, we ask you to carry out an extensive search so you can support your answer properly. 

Provide justification for above choice in atleast 50 words or more. Support your justification with scientific proofs and URLs.

\subsection{Create}\label{sec:CreateQuestion}
Your niece is 7 years old and was recently diagnosed with early Vitamin D deficiency. You want to help your aunt by providing a food plan which will help your niece to recover soon and take less supplements. The doctor has prescribed her to take 2000 I.U.(International Units) of vitamin D and he has advised to consume 2000 I.U. more through her diet. 
Create a food plan such that she receives vitamin D equivalent to 2000 I.U. as well as ALL other required nutrients.
Note: 2000 I.U. is equivalent to 50 micrograms (mcg) {50mcg = 0.05mg}

Enter data in the following format: 

'food item' - 'quantity of food' - 'unit' - 'nutrients received' - 'amount of nutrient received'

\addtocontents{toc}{\vspace{2em}} 

\backmatter


\label{Bibliography}

\lhead{\emph{Bibliography}} 

\bibliographystyle{apalike} 

\bibliography{Bibliography} 

\end{document}